\documentclass[%
 reprint,
 amsmath,amssymb,
 aps,
]{revtex4-1}

\pdfoutput=1 

\usepackage[T1]{fontenc} 

\def\jhep{JHEP}%
\def\jcap{JCAP}%
\def\prd{Phys.\ Rev.\ D}%
\def\nphysb{Nucl.\ Phys.\ B}%
\usepackage{graphicx}
\usepackage{dcolumn}
\usepackage{bm}

\newcommand{\MSBar}{\overline{MS}}
\newcommand{\MSbar}{$\MSBar$ }

\newcommand{\MEV}{ {\rm MeV} }
\newcommand{\GEV}{ {\rm GeV} }

\usepackage{booktabs}

\usepackage{rotating}
\usepackage{enumitem}

\usepackage{listings}

\usepackage{colortbl}

\usepackage{siunitx}
\usepackage[dvipsnames]{xcolor}
\usepackage{booktabs,colortbl, array}
\usepackage{pgfplots,pgfplotstable,filecontents}

\definecolor{darkgreen}{rgb}{0,0.6,0}

\graphicspath{ {Figures/} {Figures_old/}}
\usepackage{pgfplots,pgfplotstable,filecontents}

\definecolor{rulecolor}{RGB}{0,71,171}
\definecolor{tableheadcolor}{gray}{0.92}
\newcommand{\topline}{ %
        \arrayrulecolor{rulecolor}\specialrule{0.1em}{\abovetopsep}{0pt}%
        \arrayrulecolor{tableheadcolor}\specialrule{\belowrulesep}{0pt}{0pt}%
        \arrayrulecolor{rulecolor}}
\newcommand{\midtopline}{ %
        \arrayrulecolor{tableheadcolor}\specialrule{\aboverulesep}{0pt}{0pt}%
        \arrayrulecolor{rulecolor}\specialrule{\lightrulewidth}{0pt}{0pt}%
        \arrayrulecolor{white}\specialrule{\belowrulesep}{0pt}{0pt}%
        \arrayrulecolor{rulecolor}}
\newcommand{\bottomline}{ %
        \arrayrulecolor{white}\specialrule{\aboverulesep}{0pt}{0pt}%
        \arrayrulecolor{rulecolor} %
        \specialrule{\heavyrulewidth}{0pt}{\belowbottomsep}}%

\pgfplotstableset{normal/.style ={%
        header=true,
        string type,
        font=\addfontfeature{Numbers={Monospaced}}\small,
        column type=l,
        every odd row/.style={
            before row=
        },
        every head row/.style={
            before row={\topline\rowcolor{tableheadcolor}},
            after row={\midtopline}
        },
        every last row/.style={
            after row=\bottomline
        },
        col sep=&,
        row sep=\\
    }
}

\newcommand{\tsil}{\textsf{TSIL} }
\newcommand{\tsils}{\textsf{TSIL}}
\newcommand{\tarcer}{\textsf{TARCER} }
\newcommand{\tarcers}{\textsf{TARCER}}
\newcommand{\sarah}{\textsf{SARAH} }

\newcommand{\feynarts}{\textsf{FeynArts} }
\newcommand{\feynartss}{\textsf{FeynArts}}
\newcommand{\feyncalc}{\textsf{FeynCalc} }

\newcommand{\flexiblesusy}{\textsf{FlexibleSUSY} }
\newcommand{\flexiblesusys}{\textsf{FlexibleSUSY}}
\newcommand{\fire}{\textsf{FIRE} }
\newcommand{\fires}{\textsf{FIRE}}

\newcommand{\spheno}{\textsf{SPheno} }

\newcommand{\softsusy}{\textsf{SOFTSUSY} }

\newcommand{\feynhiggs}{\textsf{FeynHiggs} }

\newcommand{\feynhelpers}{\textsf{FeynHelpers} }

\newcommand{\susyhd}{\textsf{SUSYHD} }

\def\prd{Phys.\ Rev.\ D}%

\def\jcap{JCAP}%

\def\A{\mathcal{A}}
\def\B{\mathcal{B}}
\def\C{C}

\newcommand{\CC}{C\nolinebreak\hspace{-.05em}\raisebox{.4ex}{\tiny\bf +}\nolinebreak\hspace{-.10em}\raisebox{.4ex}{\tiny\bf +} }

\newcommand{\mathematica}{\textsf{Mathematica} }

\usepackage{slashed}
\newcommand{\mychi}{\raisebox{0pt}[1ex][1ex]{$\chi$}}

\def\cn{\mychi^0}
\def\cp{\mychi^+}
\def\cm{\mychi^-}
\def\chipp{\mychi^{++}}
\def\cmm{\mychi^{--}}

\def\Mp{M_{\text{pole}}}

\begin{document}

\title{\boldmath Two-loop mass splittings in electroweak multiplets: winos and minimal dark matter}

\author{James McKay and}
\email{j.mckay14@imperial.ac.uk}
\author{Pat Scott}%
 \email{p.scott@imperial.ac.uk}
\affiliation{
 Department of Physics, Imperial College London, Blackett Laboratory, Prince Consort Road, London SW7 2AZ, UK
}

\begin{abstract}
The radiatively-induced splitting of masses in electroweak multiplets is relevant for both collider phenomenology and dark matter.  Precision two-loop corrections of $\mathcal{O}$(MeV) to the triplet mass splitting in the wino limit of the minimal supersymmetric standard model can affect particle lifetimes by up to $40\%$.  We improve on previous two-loop self-energy calculations for the wino model by obtaining consistent input parameters to the calculation via two-loop renormalisation-group running, and including the effect of finite light quark masses.  We also present the first two-loop calculation of the mass splitting in an electroweak fermionic quintuplet, corresponding to the viable form of minimal dark matter (MDM). We place significant constraints on the lifetimes of the charged and doubly-charged fermions in this model.  We find that the two-loop mass splittings in the MDM quintuplet are not constant in the large-mass limit, as might naively be expected from the triplet calculation.  This is due to the influence of the additional heavy fermions in loop corrections to the gauge boson propagators.
\end{abstract}
\maketitle
\flushbottom

\section{Introduction}

Dark matter as the lightest component of an electroweak multiplet remains a viable explanation for the observed relic abundance.  One feature of this type of dark matter model is the potential for a striking signature in the form of a disappearing charged track in a collider experiment.  This is due to an order $100\,$MeV radiatively-induced mass difference between the neutral multiplet component, and the heavier charged components.  The exact length of such a track is extremely sensitive to the value of this mass difference.

At the lowest order (tree level) in perturbation theory, all components of an electroweak multiplet have the same mass.  After electroweak symmetry breaking, radiative corrections from massive gauge bosons push the physical masses of the charged components slightly above that of the neutral component \cite{Hisano2005,Cheng99}.  In many phenomenological studies, a one-loop calculation of this mass splitting is sufficient to give reasonable constraints on physical observables.  However, as we will show, due to the strong dependence on the mass splitting, two-loop corrections can result in up to a 40\% change in the lifetime of a charged multiplet component, and should be included when comparing theory with experiment.

Here we compute two-loop mass splittings for two phenomenologically-relevant electroweak multiplets.  The first is the wino in the minimal model of $R$-parity conserving supersymmetry, a fermionic electroweak triplet.  We focus specifically on the scenario where the lightest supersymmetric particle (LSP) is a pure wino (neutralino), corresponding to the neutral component of the triplet, and the rest of the supersymmetric spectrum is sufficiently massive to be decoupled. In this case, the next-to-lightest supersymmetric particle (NLSP) is also a pure wino (chargino), corresponding to the charged component of the triplet.  In this limit, a wino of mass $\sim$3\,TeV would give the correct relic abundance \cite{Hisano2007,Hryczuk2011}.  This model and the radiatively-induced mass splitting have been studied extensively, including calculation of radiative corrections to the mass splitting at two-loop order \cite{Cheng1999,Feng1999,Ibe2013}.  We refine the existing calculations by treating light quarks as massive, and by using input parameters computed using a full model spectrum.  We compare to existing results based on massless light quarks and simple threshold corrections.

The second model that we consider is the minimal dark matter (MDM; \cite{Cirelli2006,Cirelli2009}) fermionic quintuplet.  In general, MDM refers to a class of dark matter models, each consisting of the SM plus a different electroweak multiplet with some minimal set of quantum numbers and charges under the SM gauge groups.  Most models in this class have been ruled out \cite{Cai2015}, although the fermionic quintuplet with zero hypercharge is still viable.  This model has a weakly-interacting massive particle, which for a mass of $\sim$9\,TeV gives the expected dark matter relic abundance \cite{Cirelli2007,Cirelli2009}.  This model is also favoured because it stabilises the electroweak vacuum \cite{Chen2012a} by increasing the running of the electroweak gauge coupling.  Although this results in the model becoming non-perturbative at a lower scale than the SM, it at least remains perturbative until only a few orders of magnitude below the Planck scale \cite{DiLuzio2015}.  The quintuplet contains neutral, charged and doubly-charged components.  This is the first two-loop calculation of the splitting between the masses of these components.

The proper lifetime of a charged component, $\tau$, which we will express in units of (mm/c), is on the order of nanoseconds to picoseconds for the models considered here.  This corresponds to disappearing track lengths on the millimetre to centimetre scale, or more precisely about $6\,$cm \cite{Ibe2013} for the wino limit of the MSSM.  This is the motivation for many disappearing-track searches \cite{2009PhLB..672..339A,2011JHEP...05..097B,1999EPJC...11....1D,2008PhLB..664..185A,2007PhLB..644..355I,1999PhRvL..83.1731F}.  Searches with the ATLAS \cite{2013PhRvD..88k2006A} and CMS \cite{2014arXiv1411.6006C} detectors have excluded wino dark matter up to masses of $270\,$GeV and $260\,$GeV respectively.  It has been estimated that a future $100\,$TeV collider could discover pure wino dark matter up to masses of $3\,$TeV \cite{2014JHEP...08..161L,2015JHEP...01..041C}.

Similar search strategies can be applied to the MDM model.  In Ref.~\cite{2015PhRvD..92e5008O}, $8\,$TeV ATLAS and CMS results were used to exclude masses below $267\,$GeV and $293\,$GeV for the neutral components of Majorana and Dirac fermionic quintuplets, respectively.  The same paper also estimated that with a $14\,$TeV high-luminosity LHC run, Majorana MDM with a mass of up to $524$\,GeV could be discovered, as could Dirac MDM with a mass of up to $599$\,GeV.

The calculation of two-loop radiative corrections is a computationally challenging task, which has been significantly simplified with the introduction of modern tools.  Even at the most rudimentary level, determining all possible topologies is non-trivial, let alone simplifying and evaluating the resulting integrals.  Fortunately, \feynarts \cite{Hahn2001}, \feyncalc \cite{Mertig1991,Shtabovenko2016}, \tarcer \cite{Mertig1998}, \fire \cite{Smirnov2015}, \feynhelpers \cite{SHTABOVENKO201748} and \tsil \cite{Martin2006} have made each step of this process far more achievable than in the past.

The computational difficulty of the two-loop mass calculation is significantly greater for the MDM quintuplet model than for a triplet, due to the $\sim$300 additional amplitudes that must be considered compared to the triplet.  We overcome this by using a new computational framework that is almost completely automated.  This framework eventually makes the generalisation from a triplet to quintuplet trivial, and in future can be extended to make two-loop calculations achievable with even more diagrams.

Although precision two-loop self-energy corrections are essential for accurately constraining the lifetimes of charged multiplet components, the values of the input parameters used for these calculations are equally important.  Due to the scale dependence of parameters in perturbative quantum field theory, all quantities entering into a precision mass calculation are subject to potentially large uncertainties.

Computing all masses and couplings in a perturbative quantum field theory such as the MSSM is rather involved.  The physical masses must be correctly matched to corresponding running masses, which depend on the renormalisation scale.  Similarly, the couplings, which appear in the Lagrangian of the theory, are scale-dependent quantities.  Because different quantities of the calculation are defined at different scales, threshold corrections must be applied to match some low-energy theory, such as QCD, to the high energy theory of interest, such as the MSSM.  In our example, input parameters such as the running masses of the light quarks and leptons are defined in the low-energy effective QCD theory, but we are interested in determining the values of running parameters at some higher scale $Q$, so that we can use them as inputs to our two-loop self-energy calculations for the electroweak multiplet components.  To achieve this, it is necessary to numerically solve a set of ordinary differential renormalisation-group equations (RGEs) with boundary conditions defined across a hierarchy of scales, and perform the appropriate matching.

Spectrum generators are software packages that are designed to do all this in a consistent and precise way.  A number exist for the MSSM \cite{Porod:2011nf,2013CoPhC.184..899C,1993pcas.workQ...2B,2007CoPhC.176..426D,2003CoPhC.153..275P,Allanach2002}.  There are also packages intended to compute precision masses for specific states, such as \feynhiggs \cite{2000CoPhC.124...76H} and \susyhd \cite{arXiv:1504.05200}, which compute Higgs masses.  However, these packages are hardcoded to a specific model, and a specific parameterisation of that model.  In this paper, we consider both a specific limit of the MSSM, and a non-supersymmetric theory.  We therefore use tools that can create a spectrum generator from a Lagrangian, providing a consistent approach across both models.  A major part of computing a spectrum is obtaining the analytical forms of the RGE equations and the radiatively-corrected masses, threshold and tadpole corrections.  It is then the part of the spectrum generator to use numerical techniques to solve and evaluate those functions.  We use \sarah  \cite{Staub:2009bi,Staub:2010jh,Staub:2012pb,Staub2014} to produce two-loop RGEs and one-loop masses and threshold corrections, and then use \flexiblesusys \cite{Athron2015, Athron2017} to generate a spectrum generator for the MDM and MSSM models.  We link the spectrum generator to our self-energy calculations, in order to provide precision running masses and couplings as inputs to our two-loop mass-splitting calculations.

In Section \ref{sec2} of this paper, we define and detail the models that we investigate.  We then describe our calculation methods in Section \ref{sec3}, our results in Section \ref{sec4}, and summarise in Section \ref{sec5}.  We give explicit expressions for the one-loop self energies and counter-term couplings required for computing two-loop mass splittings in Appendix \ref{sec:self_energies}.

\section{Models and parameters}
\label{sec2}

\subsection{The wino limit of the MSSM}

In the wino limit of the MSSM, all supersymmetric particles except the lightest neutralino and chargino decouple from physics at the weak scale.  This corresponds to $M_2 \ll M_1, M_3, \mu$.  This implies that $gv \ll M_1,\mu$, making the lightest neutralino and chargino mass eigenstates pure winos.  Together, they constitute an $SU(2)_L$ triplet $\mychi$ with hypercharge $Y=0$, coupled to the SM via the electroweak sector.  The \MSbar renormalised Lagrangian is
\begin{align}
\mathcal{L}=\mathcal{L}_{\text{SM}}+\frac{1}{2}\overline{\mychi}\,(i\slashed{\mathcal{D}}-\hat{M})\,\mychi \label{eqn:wino_limit}
\end{align}
where $\mathcal{L}_{\text{SM}}$ is the SM Lagrangian, $\hat{M}$ is the degenerate \MSbar tree-level mass of the triplet and $\slashed{\mathcal{D}}$ is the $SU(2)_L$ covariant derivative.  Expanding out the covariant derivative gives
\begin{align}
\begin{split}
\mathcal{L}=&\mathcal{L}_{\text{SM}}+\frac{1}{2}\overline{\mychi^0}(i\slashed{\partial}-M)\mychi^0+\frac{1}{2}\overline{\mychi^+}(i\slashed{\partial}-M)\mychi^{+}\\
&+g\left (\overline{\mychi^+}\gamma_{\mu}\mychi^+ \right)\left(s_wA_{\mu}+c_wZ_{\mu}\right)+g\left(\overline{\mychi^+}\gamma_{\mu}\mychi^0\right)W_{\mu}^++\text{h.c.}
\end{split}
\end{align}
The triplet therefore couples to the SM via the electroweak gauge bosons only.  It is clear that at tree level, the charged and neutral components have the same mass, $\hat{M}$.  We will express the physical masses of the neutral and charged components as $\Mp^0$ and $\Mp^+$ respectively.

When implementing the electroweak triplet in \sarah model files, we express it in matrix form:
\begin{align}
\mychi = \left( \begin{array}{cc} \mychi^0 /\sqrt{2} &  \mychi^+ \\
\mychi^- & -\mychi^0/\sqrt{2}
\end{array}\right).
\end{align}

\subsection{The electroweak quintuplet}

In the remaining viable version of MDM, a fermionic $SU(2)_L$ quintuplet $\mychi$ with hyper-charge $Y=0$ is coupled to the SM via the $SU(2)_L$ gauge sector.  Analogous to the triplet case, the Lagrangian is
\begin{align}
\mathcal{L}=\mathcal{L}_{\text{SM}}+\frac{1}{2}\overline{\mychi}\,(i\slashed{\mathcal{D}}-\hat{M})\,\mychi,
\end{align}
where $\slashed{\mathcal{D}}$ is the $SU(2)_L$ covariant derivative and $\hat{M}$ is the \MSbar renormalised tree-level quintuplet mass.  Expanding the $SU(2)_L$ covariant derivative $\slashed{\mathcal{D}}$, we obtain the interaction terms
\begin{align}
\begin{split}
\mathcal{L}=&\mathcal{L}_{\text{SM}}+\frac{1}{2}\overline{\mychi^0}(i\slashed{\partial}-M)\mychi^0+\frac{1}{2}\overline{\mychi^+}(i\slashed{\partial}-M)\mychi^{+}\\
&+\frac{1}{2}\overline{\mychi^{++}}(i\slashed{\partial}-M)\mychi^{++}\\
&+g\left(\overline{\mychi^+}\gamma_{\mu}\mychi^++2\overline{\mychi^{++}}\gamma_{\mu}\mychi^{++}\right)\left(s_wA_{\mu}+c_wZ_{\mu}\right)\\
&+g\left(\sqrt{3}\,\overline{\mychi^+}\gamma_{\mu}\mychi^0+\sqrt{2}\,\overline{\mychi^{++}}\gamma_{\mu}\mychi^+\right)W_{\mu}^++\text{h.c.}
\end{split}
\end{align}
As with the triplet, the quintuplet couples only to the photon and $W$ and $Z$ bosons at tree level.  We will express the physical masses of the neutral, charged and doubly-charged components as $\Mp^0$, $\Mp^+$ and $\Mp^{++}$ respectively.

For implementing the quintuplet $\chi$ in a \sarah model file, we express it in tensor representation as \cite{2012PhRvD..86a3006K}
\begin{align}
\begin{split}
\mychi_{1111} &=  \chipp \ \ , \ \ \mychi_{1112} = \frac{1}{\sqrt{4}} \cp \ \ , \ \ \mychi_{1122} = \frac{1}{\sqrt{6}}\cn \\
\mychi_{1222} &= -\frac{1}{\sqrt{4}}  \cm \ \ , \ \ \mychi_{2222} =  \cmm
\end{split}
\end{align}
where the relative signs are chosen such that $\mychi$ is isospin self-conjugate \cite{2015JCAP...10..058G}.  In this representation the mass term is given by
\begin{equation}
\overline{\mychi^C}\mychi\equiv\overline{\mychi^C}_{ijkl}\mychi_{i'j'k'l'}\epsilon^{ii'}\epsilon^{jj'}\epsilon^{kk'}\epsilon^{ll'}.
\end{equation}

\subsection{Input parameters}

In this paper, we use a fully-computed model spectrum to obtain the input parameters for our self-energy calculations.  To generate the spectrum, we therefore require a full set of SM input parameters.  These are given in Table \ref{table:params}.  The central values and experimental uncertainties are from the latest Particle Data Group tables \cite{Patrignani2016}.  We quantify the parametric sensitivity of the mass splitting to each of these uncertainties by varying one parameter at a time, and holding the rest fixed.  We show the results of this exercise in Table \ref{tab:uncertainty_analysis}, at a phenomenologically relevant value of the degenerate mass for each model.

The renormalisation scale $Q$ is an important input parameter in our calculation.  This is the scale to which all mass parameters and couplings are run, and where the self energies, and subsequent pole mass, are evaluated.  The range of this parameter should reflect the scale of missing logarithmic corrections in the calculation, which are of the form $\log(m/Q)$ for some mass $m$.  When using a non-iterative method for computing the multiplet mass splitting, we find that the dominant missing logarithmic corrections come from masses near the electroweak scale.  Contributions from the multiplet itself, with masses around the TeV scale, are cancelled.  See Ref.\ \cite{2017arXiv171001511M} for a detailed discussion.  Therefore, for this study it is sufficient to vary the renormalisation scale around the mass of the top quark.  We therefore choose the range $m_t/2\leq Q \leq 2m_t$.

\begin{table}[tp]
\centering
\caption{Input parameters and uncertainties used for the calculations in this study (unless stated otherwise).  These ranges and central values are taken from the latest Particle Data Group tables \cite{Patrignani2016}.} \label{table:params}
\vspace{0.4cm}
\begin{tabular}{l@{\hspace{4mm}}c@{}c}
\hline
Parameter & & Values \\
\hline
Electromagnetic coupling & $1/\alpha^{\overline{MS}}(m_Z)$        & $127.940(42)$       \\
Top pole mass  & \phantom{$^{\overline{MS}}$}$m_t$\phantom{$^{\overline{MS}}$}  &  $173.34(2.28)$\,GeV\\
Higgs pole mass & \phantom{$^{\overline{MS}}$}$m_h$\phantom{$^{\overline{MS}}$}  & $125.5(1.6)$\,GeV \\
W pole mass & \phantom{$^{\overline{MS}}$}$m_W$\phantom{$^{\overline{MS}}$}  & $80.385(15)$\,GeV \\
Z pole mass & \phantom{$^{\overline{MS}}$}$m_Z$\phantom{$^{\overline{MS}}$}  & $91.1876(21)$\,GeV \\
Electron pole mass & \phantom{$^{\overline{MS}}$}$m_e$\phantom{$^{\overline{MS}}$}  & $0.510 998 9461(31)$\,MeV \\
Muon pole mass & \phantom{$^{\overline{MS}}$}$m_{\mu}$\phantom{$^{\overline{MS}}$}  & $ 105.6583745(24)$\,MeV \\
Tau pole mass & \phantom{$^{\overline{MS}}$}$m_{\tau}$\phantom{$^{\overline{MS}}$}  & $ 1776.86(12)$\,MeV \\
Down quark mass & $m_d^{\overline{MS}}(2\,\text{GeV})$  &   $4.80(96)$\,MeV  \\
Up quark mass & $m_u^{\overline{MS}}(2\,\text{GeV})$     & $2.30(46)$\,MeV \\
Strange quark mass & $m_s^{\overline{MS}}(2\,\text{GeV})$  & $95(15)$\,MeV\\
Charm quark mass & $m_c^{\overline{MS}}(m_c)$ & $1.275(75)$\,GeV \\
Bottom quark mass & $m_b^{\overline{MS}}(m_b)$    & $4.18(9)$\,GeV \\
Strong coupling & $\alpha_S^{\overline{MS}}(m_Z)$  & $0.1181(11)$\\
Renormalisation scale & $Q$    & $m_t/2 - 2m_t$\\
\hline\end{tabular}
\end{table}

\section{Method}
\label{sec3}

To determine the mass splitting we must compute the physical, or \textit{pole}, masses of the multiplet components to a fixed order in perturbation theory.  The definition of a pole mass is the complex pole of the two-point propagator, which for a fermion has a denominator given by the one-particle irreducible effective two-point function
\begin{equation}
\Gamma_2=\slashed{p}-\hat{M}+\Sigma_K(p^2)\slashed{p}+\Sigma_M(p^2).
\label{eqn:propagator}
\end{equation}
Here $p_{\mu}$ is the four-momentum of the particle, $\hat{M}$ is the tree-level \MSbar mass and $\slashed{p}=\gamma^{\mu}p_{\mu}$.  The self energy, $\Sigma(p^2)=\Sigma_M(p^2)+\slashed{p}\Sigma_K(p^2)$, is in general a function of the renormalisation scale and any relevant masses or couplings in the theory.  We will expand the self energy up to second order in a perturbation parameter $\alpha$, for a two-loop result.

The pole mass is obtained by demanding $\Gamma_2=0$.  This can be achieved by setting $p^2=\Mp^2$ (and $\slashed{p}=\Mp$), and solving the resulting implicit expression for the pole mass
\begin{align}
\Mp=\text{Re}\left[\frac{\hat{M}-\Sigma_M(\Mp^2)}{1+\Sigma_K(\Mp^2)}\right] \label{eqn:M_pole_iterative}.
\end{align}

Solving (\ref{eqn:M_pole_iterative}) iteratively results in unwanted scale-dependent logarithms.  Alternatively, one can take advantage of the perturbative nature of this expression to write down an explicit result for the pole mass that preserves a fortunate cancellation of the scale-dependent logarithms.  Working to the two-loop order, this expression is
\begin{align}
\begin{split}
&\Mp=\left[\hat{M}-\Sigma^{(1)}_M-\Sigma^{(2)}_M-\hat{M}\Sigma^{(1)}_K-\hat{M}\Sigma^{(2)}_K\right.\\
&+(\Sigma^{(1)}_M+\hat{M}\Sigma^{(1)}_K)(\Sigma^{(1)}_K+2\hat{M}\dot{\Sigma}^{(1)}_M+2\hat{M}^2\dot{\Sigma}^{(1)}_K)\\
&\left.+\mathcal{O}\left(\alpha^3\right)\right]_{p^2=\hat{M}^2}, \label{eqn:M_pole_explicit}
\end{split}
\end{align}
where $\Sigma^{(n)}_X=\Sigma^{(n)}_X(p^2)$.  For a full review of this method and the implications of the iterative procedure see Ref.\ \cite{2017arXiv171001511M}.

For this study we use the Feynman-'t Hooft ($\xi=1$) gauge for all calculations.  One-loop mass splittings computed in the Landau ($\xi=0$), Feynman-'t Hooft ($\xi=1$) and Fried-Yennie ($\xi=3$) gauges can also be found in Ref.\ \cite{2017arXiv171001511M}.

\tikzset{
    photon/.style={decorate, decoration={snake,segment length=2mm, amplitude=0.8mm}, draw=black},
    wino/.style={draw=black},
    scalar/.style={draw=black,style={dash pattern=on 4pt off 4pt}},
    ghost/.style={draw=black,style={dash pattern=on 1pt off 1pt}}
}

\tikzstyle{vertex} = [circle,fill=black,inner sep=0pt,minimum size=3pt]
\tikzstyle{empty} = [circle,fill=none,inner sep=0pt,minimum size=3pt]
\tikzstyle{SMcorrection} = [circle,fill=black,inner sep=0pt,minimum size=15pt]
\tikzstyle{counterterm} = [circle,fill=black,inner sep=0pt,minimum size=10pt]
\newcommand{\CIRCLE}{$\mathbin{\tikz [x=2.28ex,y=2.28ex,line width=.5ex, black] \draw (0,0) -- (1,1) (0,1) -- (1,0);}$}%
\newcommand{\Cross}{$\mathbin{\tikz [x=2.28ex,y=2.28ex,line width=.5ex, gray] \draw (0,0) -- (1,1) (0,1) -- (1,0);}$}%
\newcommand{\CT}{$\mathbin{\tikz [x=1.28ex,y=1.28ex,line width=.5ex, gray] \draw (0,0) -- (1,1) (0,1) -- (1,0);}$}%

\begin{figure*}[t!]
\center{\includegraphics{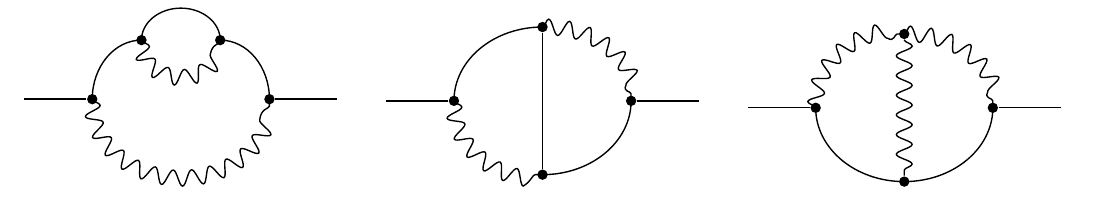}}
\caption{Two-loop diagrams involving only the gauge bosons and multiplet fermions.  Solid lines indicate multiplet fermions ($\chi^{0},\, \chi^{\pm},\, \chi^{\pm\pm}$) and wiggly lines electroweak vector bosons ($W^{\pm}$, $Z$, $\gamma$).} \label{fig:Feynman_diagrams}
\end{figure*}

\begin{figure*}[t!]
\center{\includegraphics{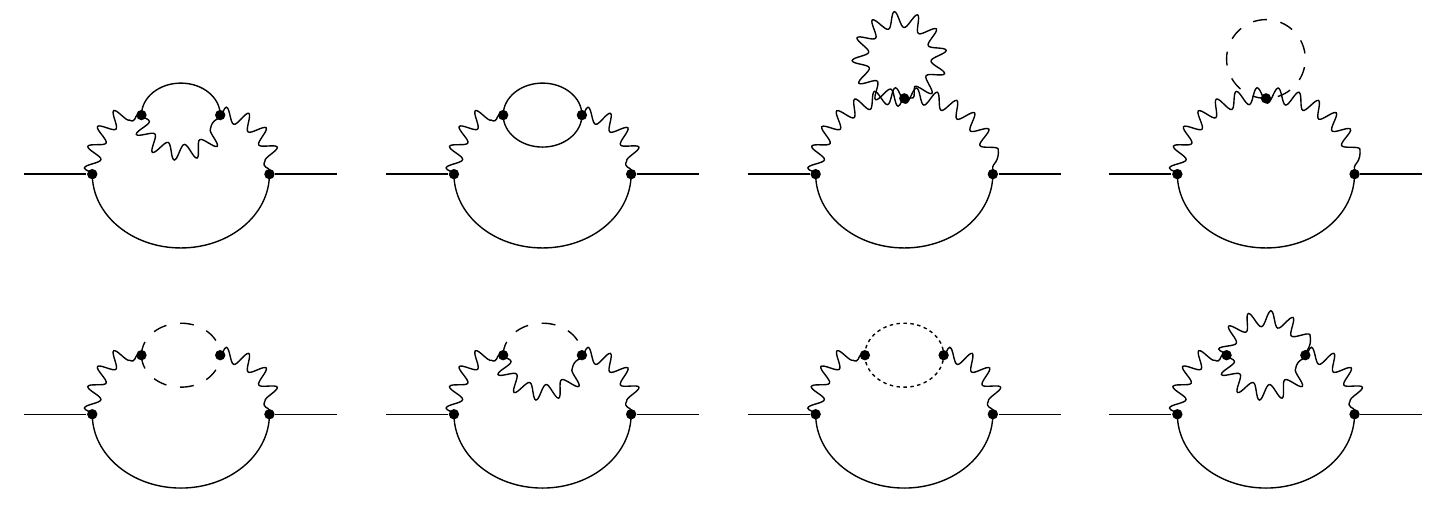}}
\caption{Two-loop diagrams formed by reinserting the 1-loop gauge boson self-energy into its own propagator.  Solid lines indicate fermions ($\chi^{0},\, \chi^{\pm},\, \chi^{\pm\pm},\, q,\, l,\, \nu$), wiggly lines electroweak vector bosons ($W^{\pm}$, $Z$, $\gamma$), dashed lines scalars (Higgs and Goldstone bosons) and dotted lines indicate ghosts.} \label{fig:Feynman_diagrams2}
\end{figure*}

\begin{figure*}[t!]
\center{\includegraphics{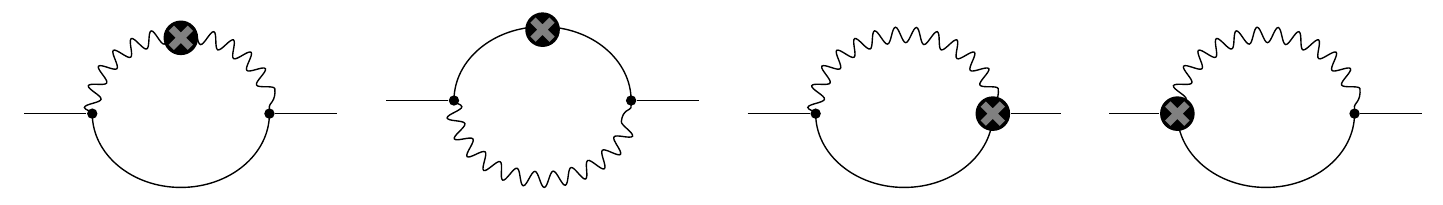}}
\caption{Two-loop counter-term diagrams.  Small circles with crosses indicate counter-term insertions.  Solid lines indicate multiplet fermions ($\chi^{0},\, \chi^{\pm},\, \chi^{\pm\pm}$) and wiggly lines electroweak vector bosons ($W^{\pm}$, $Z$, $\gamma$).} \label{fig:Feynman_diagramsCT}
\end{figure*}

In the wino limit of the MSSM there are about 200 two-loop diagrams, and about 500 for the MDM quintuplet model.  The generic two-loop topologies are given in Figures \ref{fig:Feynman_diagrams} and \ref{fig:Feynman_diagrams2}, and counter-term diagrams of two-loop order in Figure \ref{fig:Feynman_diagramsCT}.  We determine the counter-term couplings from the one-loop self energies of the electroweak gauge bosons and electroweak multiplets.

\subsection{Details of self-energy calculation}\label{sec:calc_method}

In this subsection, we describe our automated process for calculating self-energies at two loops.

A complete self-energy calculation (at any order) requires the construction of a symbolic amplitude, followed by its numerical evaluation.  In general, interfaces between tools are sufficient for generating symbolic amplitudes at both one and two-loop level.  For one-loop calculations, the evaluation step can be performed with various existing tools: \feynhelpers \cite{Shtabovenko2016} provides analytic one-loop amplitudes for this purpose, and other codes do this by making use of the \textsf{LoopTools} package \cite{Romao2006} (e.g.\ \sarah \cite{Staub2014} interfaced to either \spheno \cite{Porod2003} or \flexiblesusy \cite{Athron2015, Athron2017}).

The interface between the tools available for generic two-loop calculations is only complete up to the stage of the symbolic amplitude.  The necessary conversions exist between \feynartss, \feyncalc and \tarcers, but the final step of numerical evaluation requires significant user intervention.  The \tsil library provides numerical, and in some cases analytical, solutions for the basis integrals that appear in two-loop self-energies.  However, in order to make use of these, one must construct a \CC interface to call the \tsil libraries and then use them to evaluate the amplitudes.  Although the \tsil functions are extremely user-friendly, making use of them from a symbolic \mathematica expression provided by one of the other tools is highly non-trivial. There is therefore no automated method for obtaining numerical implementations of two-loop amplitudes.  Given that there can be hundreds or even thousands of such amplitudes, this makes the final step of the calculation an arduous process.  By completely automating the generation of this \CC interface with a new software framework, we have been able to dramatically simplify the process of computing two-loop self energies.  This framework has already been used to generate two-loop amplitudes used in Ref.\ \cite{2017arXiv171001511M}.

Our method also makes it possible to split the calculation of many loop diagrams into manageable pieces.  Simultaneously computing $\mathcal{O}(10)$ different amplitudes (of distinctly different masses and/or topologies) with symbolic tools like \feyncalc takes an extremely long time, as \feyncalc attempts to symbolically simplify the amplitudes.  On the other hand, keeping track of all terms on a diagram-by-diagram basis is a serious task by any manual or even semi-automated method.  By completely automating the whole process, we are instead able to keep track of all terms, and simply evaluate them independently and numerically.  On a modest computing setup, this is the only way to obtain a result in a feasible timeframe without additional user intervention.

We calculate the amplitudes either one diagram at a time, or in selected groups, using \feynartss, \feyncalc and \fires, run from \CC via the Wolfram Symbolic Transfer Protocol (WSTP).  We decompose the resultant symbolic amplitudes into lists of coefficients to be applied to basis integrals, and keep a master list of all the basis integrals required.

The algorithm begins by evaluating the finite part of the amplitude $\A$.  It then computes the coefficients $\{\C_1,\C_2, \ldots  \}$ of every possible basis integral $\{\B_1,\B_2,\ldots \}$.  For the non-zero $\C_i$, it then constructs a trial amplitude of the form
\begin{equation}
\A_{\rm trial} = \C_1 \B_1 + \C_2 \B_2 + \ldots
\end{equation}
and checks the difference $\A - \A_{trial}$ for the presence of basis integrals with non-zero coefficients, in order to identify any cross-terms that have been double-counted in the first step.  From the set of basis integrals $\{\B_i,\B_j,\ldots \}$ with non-zero coefficients at this stage, the algorithm then creates new `compound basis integrals' $\B_{ij} =  \B_i\B_j$, and presents them to \mathematica as unified objects.  We can then instruct \mathematica to extract new coefficients $\C_{ij}$ for the compound basis integrals.  The final amplitude is then
\begin{align*}
 \A_{trial}\ = \ & \ \ \C_1 \B_1 + \C_2 \B_2 + \ldots \\
  &-\frac{1}{2} \C_{12} (\B_1\B_2)  -\frac{1}{2} \C_{21} (\B_2\B_1) - \ldots \\
  & + C_{11} (\B_1\B_1) + C_{22}(\B_2\B_2) + \ldots
\end{align*}
where $ \C_{ij}$ is the coefficient of $\B_i\B_j$ in the original amplitude $\A$.  We convert these coefficients into \CC format, and generate numerical routines for evaluating both them and the relevant basis integrals.

This automated framework is fully generic, allowing numerical routines to be generated for two-loop diagrams in almost any \feynarts model file.  The only limitations are computational: problems involving over $\sim$1000 diagrams require long runtimes to generate the amplitudes, and produce large amounts of generated code.  Other features include: \begin{itemize}
\item automatic determination of one-loop counter-term couplings for two-point diagrams (using the one-loop self-energies),
\item optimisation of the evaluation of the two-loop basis integrals, by automatically determining which integrals can be evaluated in symmetry groups, and
\item flexibility and reusability of precomputed amplitudes (by separating the symbolic calculations from the final code generation).
\end{itemize}
We intend to make an open-source release of the full package in the near future.

In this paper, we use \feyncalc \textsf{9.2.0} \cite{Mertig1991,Shtabovenko2016} and \feynarts \textsf{3.9} \cite{Hahn2001} to obtain symbolic amplitudes, and reduce them to basis integrals with \fire \textsf{5} \cite{Smirnov2015} (via \feynhelpers \textsf{1.0.0} \cite{SHTABOVENKO201748}) and \tarcer \textsf{2.0} \cite{Mertig1998}.  We evaluate the basis integrals using \tsil \textsf{1.41} \cite{Martin2006} and analytical forms from the literature \cite{Pierce1997}.

\subsection{Check for divergence free-result}

It is important to confirm that the pole masses are free of non-physical divergences.  Ultra-violet (UV) divergences can be regulated using dimensional regularisation by computing in $D=4-2\epsilon$ dimensions and using modified minimal subtraction.  Using the generated symbolic amplitudes and our numerical implementation, we have confirmed that the individual pole masses are free from any poles in $\epsilon$ when the appropriate counter-terms are included.

Infra-red divergences arise from the zero mass of the photon.  To regulate these divergences, we retain an explicit mass $m_{\gamma}$ for the photon throughout the calculation, and take the limit $m_{\gamma}\rightarrow0$ in the evaluation.  IR-divergent diagrams exist at two-loop order, but their divergences are cancelled by the derivative of the one-loop self-energies in the two-loop expansion of the pole mass (Eq.\ \ref{eqn:M_pole_explicit}).  The proof of this cancellation is given in Ref.~\cite{Ibe2013} for the wino model.  The analogous result holds identically for the MDM quintuplet, so we do not repeat the details here.

We also encounter `fictitious' IR divergences in our numerical implementation.  These can arise from including a finite photon mass when attempting to evaluate non-IR divergent diagrams.  We nonetheless include this mass for all diagrams, as in some cases, taking a zero photon mass \textit{before} the tensor integral reduction causes the tensor integrals to reduce to basis integrals that are not available in current mathematical libraries. Using a regulator mass enables the reduction to proceed further, giving a result in terms of known basis integrals.  The price to pay for this convenience is an apparent IR singularity in the result: the amplitude picks up $\mathcal{O}(1/m_{\gamma}^2)$ terms.  However, the sum of the coefficients of these terms is numerically equivalent to zero for every diagram (i.e.\ to within a small factor of the floating-point machine accuracy times the largest individual coefficient).  We therefore always see numerically that these terms cancel, even if the integral reduction fails to cancel them symbolically.  We take care in our evaluation step to explicitly check for the numerical cancellation, and to then remove the terms \textit{before} taking the limit $m_{\gamma}\rightarrow 0$, as the latter would otherwise cause numerical cancellation errors between the $\mathcal{O}(1/m_{\gamma}^2)$ terms to blow up and dominate the result.

Also, because the basis integral $T(x,y,z)$ is not defined for small $x$, in the limit of $m_{\gamma}\rightarrow0$ we make the replacement $T(x,y,z) \equiv \bar{T}(x,y,z) - B(y,z)\log(x/Q^2)$ \cite{Martin2006}, where $B$ is another basis integral.  This will cancel with other terms in the amplitude of the form $A(x)B(y,z) = x\left[\log(x/Q^2)-1\right]B(y,z)$, and because $\bar{T}(0,y,z)$ is finite, will give a total that is IR safe.

\subsection{Spectrum calculation}

We use \flexiblesusys \textsf{1.7.4}\footnote{\flexiblesusy also uses some code pieces from \softsusy \cite{Allanach2002,Allanach:2013kza}.} \cite{Athron2015, Athron2017} to create a spectrum generator, based on output from \sarah \textsf{4.8.0} \cite{Staub:2009bi,Staub:2010jh,Staub:2012pb,Staub2014}.  This provides two-loop RGEs, one-loop threshold and tadpole corrections and one-loop self energies for all fields.  Because the spectrum generator requires a tree-level parameter prior to computing the loop-corrected EWSB conditions, the Higgs pole mass is an output rather than an input parameter.  Thus we also employ a simple iterative procedure to determine the correct input value for the Lagrangian Higgs mass parameter $\mu$, such that the observed Higgs pole mass is produced.

From the computed spectrum we extract the \MSbar masses for the gauge bosons, Higgs and quarks at a common scale $Q$.  We also extract the running electroweak gauge couplings at $Q$ and use them to compute the value of $\alpha_{\text{EM}}(Q)$ from the relation
\begin{align}
\alpha_{\text{EM}} = \frac{g_1^2g_2^2}{4\pi (g_1^2+g_2^2)}.
\end{align}
From these parameters we compute the Weinberg angle $\theta_W=\arccos(m_W/m_Z)$ and the Higgs vacuum expectation value $v=2\sin(\theta_W)m_W  /  \sqrt{4\pi}$.  This preserves the required tree-level relations that are necessary to retain the proper cancellations between parts of the self-energies of the charged and neutral multiplet components.

It is important to consider threshold corrections when matching the SM to the wino or MDM model at two loops.  These corrections include the determination of \MSbar masses consistent with a specified physical pole mass (particularly important for the $W$ and $Z$ bosons), and matching the \MSbar gauge couplings in the SM to the model containing additional fermions.  The relevant threshold correction for the electroweak coupling is
\begin{align}
\alpha_{\text{EM,\,wino}}(Q) = \alpha_{\text{EM,\,SM}}(Q) \left[ 1- \frac{X\alpha_{\text{EM}}(Q)}{3\pi}\log\left(\frac{\hat{M}(Q)}{Q}\right)\right]^{-1},\label{eq:tc}
\end{align}
where $X=2$ for the wino model and $X=10$ for the MDM quintuplet.  \flexiblesusy applies this correction and does the mass matching, iteratively, at $Q=m_Z$.

For the one-loop calculation, we do not need to apply threshold corrections, as they are of the next loop order.  If these corrections are applied, then important cancellations do not occur between the threshold corrections and the self-energies, resulting in a spurious logarithmic increase or decrease in the mass splitting.   When calculating one-loop mass splittings, we therefore use pole masses in place of the \MSbar masses, and neglect the threshold corrections to the gauge couplings.  This is consistent with the method of Ref.\ \cite{Ibe2013}.

\section{Results}
\label{sec4}

\subsection{The wino limit of the MSSM}

\begin{figure*}
\centering
\includegraphics[width=0.9\textwidth]{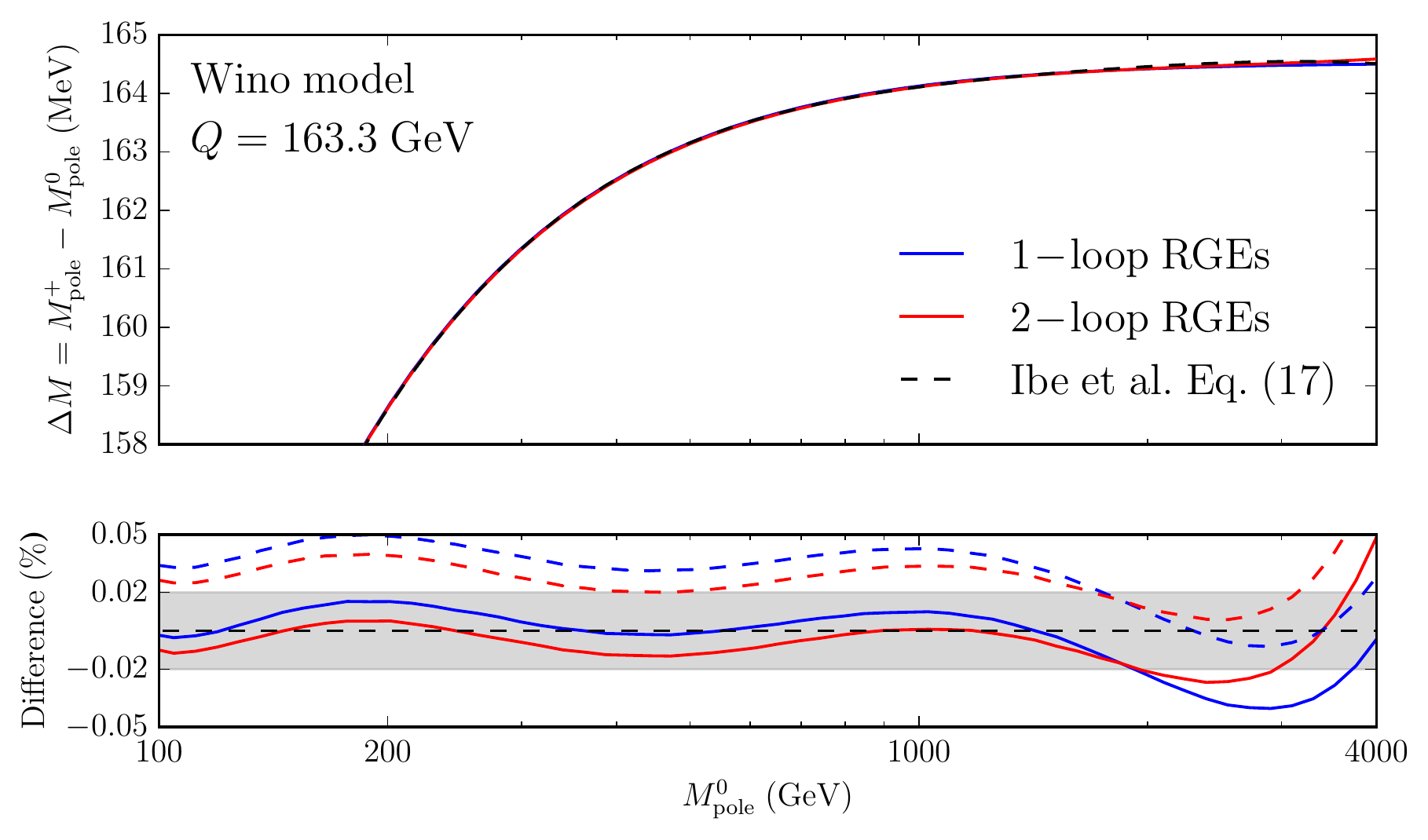}
\caption{The two-loop mass splitting in the wino model for $m_t=173.2\,$GeV, $m_h = 125.5\,$ GeV, $\alpha_S^{\overline{MS}}(m_Z)=0.1184$ and $Q=163.3\,$GeV.  In both panels the black dashed line is the fit given by Eq.\ (17) of Ref.~\cite{Ibe2013}, and the grey region in the lower panel is the stated deviation of this fit from the actual result.  The solid lines indicate our result with one and two-loop RGEs in blue and red respectively, with all light quark masses taken to be zero.  The red and blue dashed lines in the lower panel correspond again to one and two-loop RGEs respectively, but with all light quark masses included.}\label{fig:Ibe_compare}
\end{figure*}

As electroweak mass splittings have already been studied at the two-loop level in the wino limit of the MSSM \cite{Yamada2010,Ibe2013}, we are able to compare our results to the previous ones, and in the process demonstrate the impacts of the improvements that we have made in this paper.  This also serves as a validation of the consistency of our method, in particular the use of a full spectrum generator, before applying our method to the MDM quintuplet.

In Figure \ref{fig:Ibe_compare} we compare our two-loop results for the mass splitting $\Delta M \equiv \Mp^+ - \Mp^0$ to the one given by the polynomial fit in Eq.\ (17) of Ref.\ \cite{Ibe2013}.  For consistency, we use the same top pole mass $m_t=173.2\,$GeV \cite{SatoPrivate} and strong coupling $\alpha(m_Z) = 0.1184$ as in Ref.\ \cite{Ibe2013}.  We compare with both one and two-loop RGEs, and with both finite and zero masses for the light quarks.  The authors of Ref.\ \cite{Ibe2013} state that their polynomial fit gives less than a $0.02\%$ deviation from the true value over the mass range $100-4000$\,GeV, so we expect to be able to achieve a result close to this when comparing with our calculation.  We see that our equivalent result (one-loop RGEs, zero light quark masses) is in good agreement with theirs, with the deviation clearly the result of ringing from the polynomial fit rather than an inconsistency between the methods used.  This ringing is worst at large masses, where we expect the mass splitting to be constant; the polynomial fit fails to properly represent this behaviour.  Over the whole mass range we have no more than a $0.05$\% deviation from this previous calculation.  The impact of the light quark masses is to increase the mass difference by about 0.03--0.04\% across the whole mass range.

The consistency of our result with that of Ref.\ \cite{Ibe2013} can also be seen by making a polynomial fit to the curve computed with 1-loop RGEs, massless light quarks, $Q=163.3\,$GeV, $m_t=173.2\,$GeV and $\alpha(m_Z) = 0.1184$, over the range $100\,\mathrm{GeV}\le M^0_\mathrm{pole} \le 4$\,TeV:
\begin{widetext}
\begin{eqnarray}
\frac{\Delta M}{1~\MEV} &=&  - 412.2
+304.7 \left(\ln \frac{\Mp^0}{1~\GEV}\right)
-60.71 \left(\ln\frac{\Mp^0}{1~\GEV}\right)^2
+5.403 \left(\ln\frac{\Mp^0}{1~\GEV}\right)^3\label{eqn:fit1}
-0.181 \left(\ln\frac{\Mp^0}{1~\GEV}\right)^4.
\end{eqnarray}
\end{widetext}
This is in very close agreement with Eq.\ (17) of Ref.\ \cite{Ibe2013}.

\begin{figure*}
\centering
\includegraphics[width=0.5\textwidth]{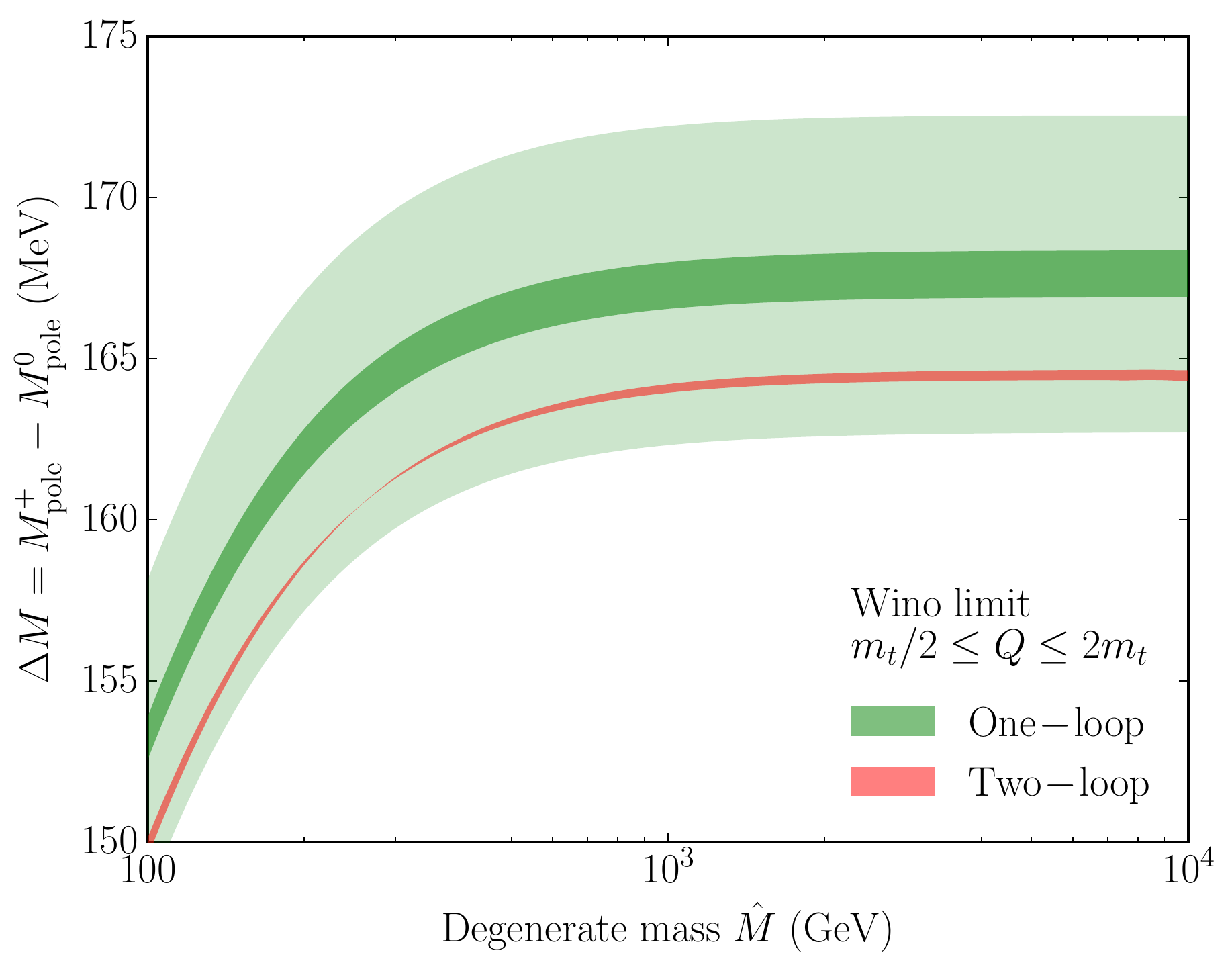}\includegraphics[width=0.5\textwidth]{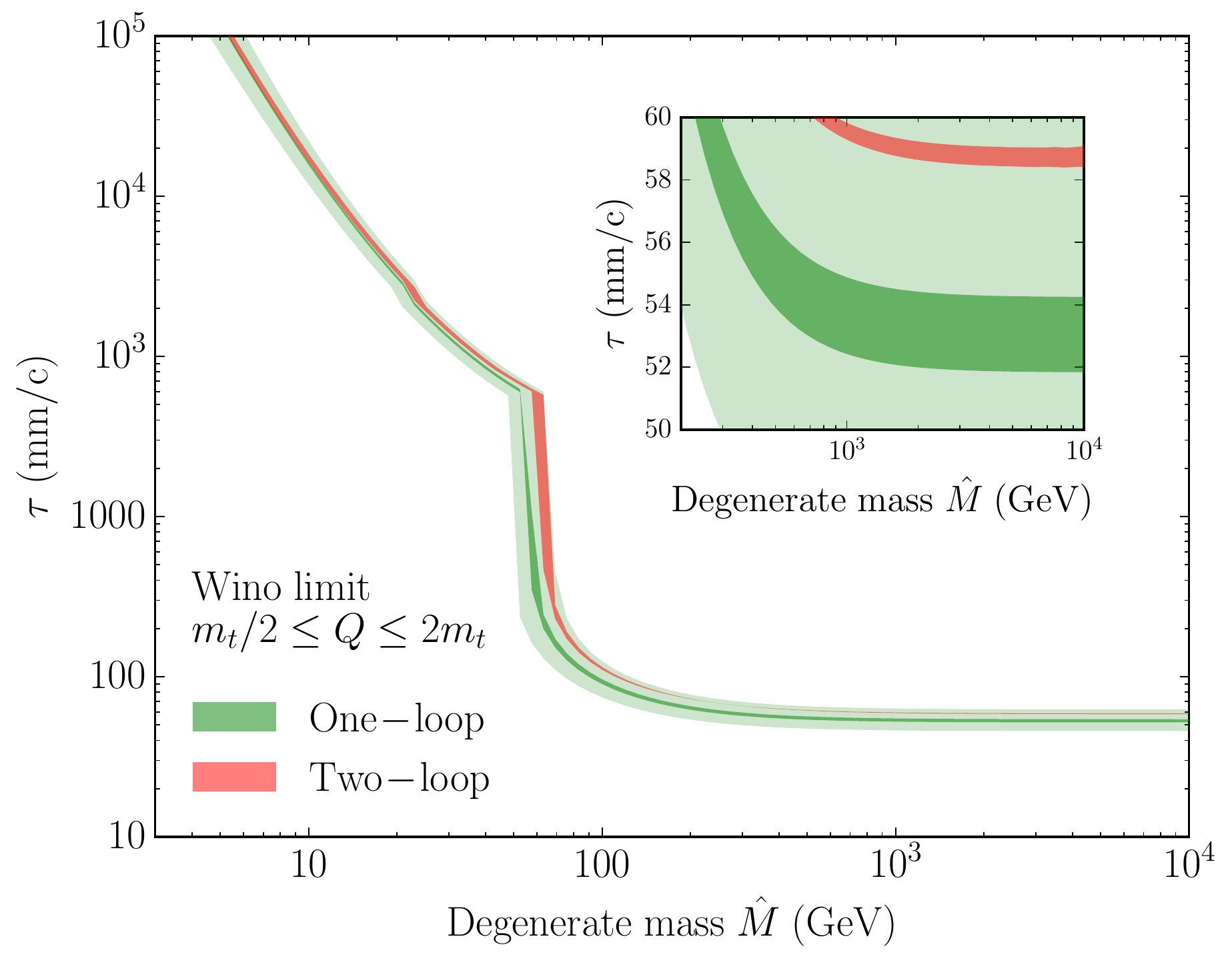}
\caption{The two-loop mass splitting (\textit{left}) and decay lifetime of the chargino (\textit{right}) in the wino model as a function of the degenerate tree-level \MSbar mass.  The green and red bands are the respective ranges of the one and two-loop mass splittings when $Q$ is varied continuously between $m_t/2$ and $2m_t$.  The light green band is the estimated uncertainty on the one-loop result using Eq.~(\ref{eqn:error_estimate}).}\label{fig:wino_two_loop}
\end{figure*}

\begin{figure*}
\centering
\includegraphics[width=0.5\textwidth]{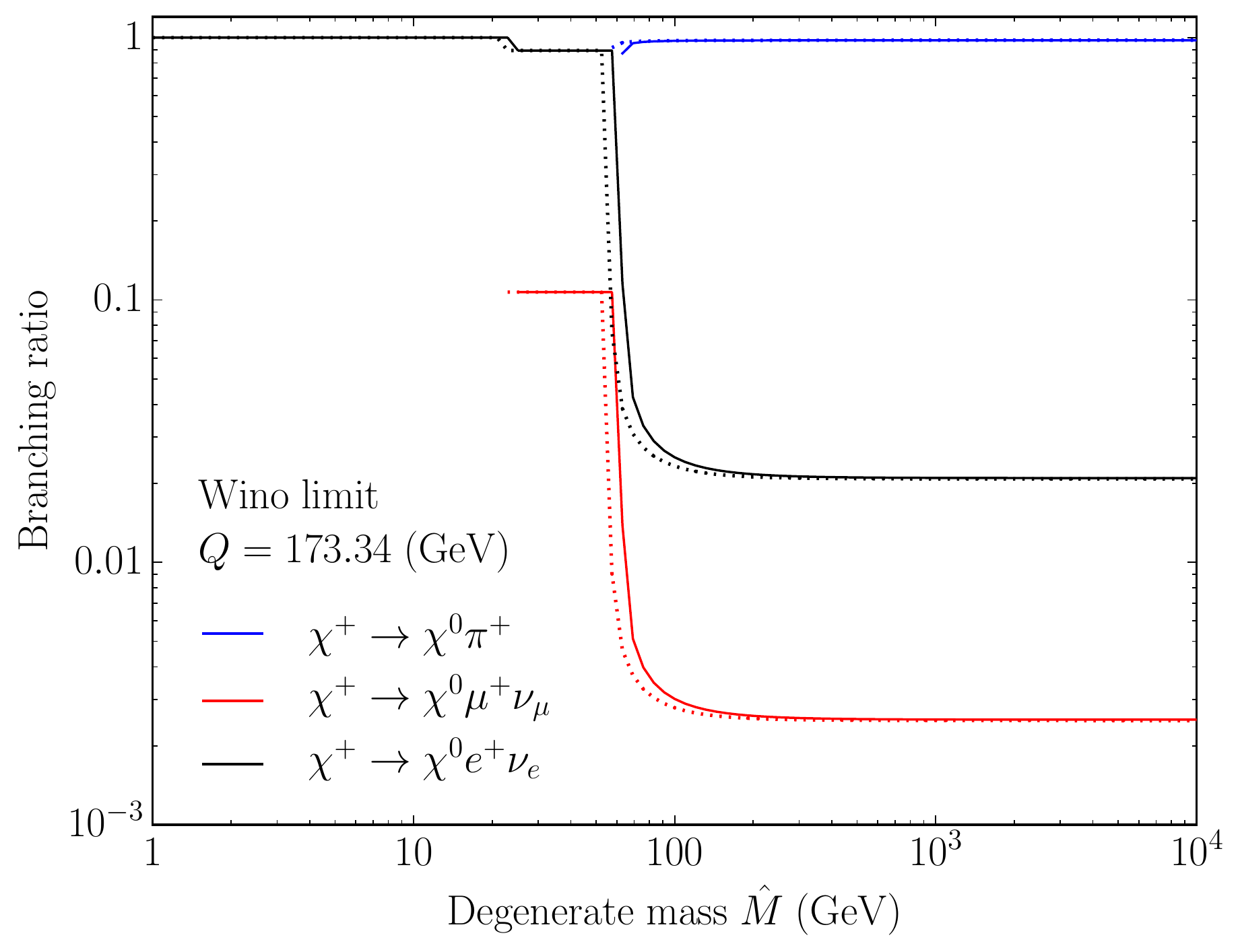}\
\caption{Branching fractions for $\chi^+\rightarrow X\chi^0$ in the wino limit of the MSSM, where $X \in \{e\nu_e,\mu\nu_\mu,\pi\}$.  Solid lines are the branching fractions using the two-loop mass splitting, and dotted use the one-loop result, both evaluated at $Q=m_t=173.34\,$GeV.}\label{fig:BFs}
\end{figure*}

In Figure \ref{fig:wino_two_loop} we present the two-loop mass splitting in the wino model using the parameters in Table \ref{table:params}, two-loop RGEs and non-zero light quark masses.  The dark green and red uncertainty bands are given by the maximum and minimum $\Delta M$ possible for values of $Q$ between $m_t/2$ and $2m_t$.  For the two-loop mass splitting, at values of $\hat{M}\lesssim m_t$ the minimum splitting occurs at $Q=m_t/2$, whereas for $\hat{M} \gtrsim 2m_t$ the minimum occurs at $Q=2m_t$.  For intermediate values $\hat{M}\sim m_t$, around the point where the crossover occurs, we find that the extrema occur at values of $Q$ inside the chosen range.  As a result, although the uncertainty band appears by eye to become very narrow, it does in fact maintain a non-zero width.  At even lower multiplet masses than shown here ($\hat{M}\lesssim100$\,GeV), the two-loop uncertainty from scale variation on the mass splitting is comparable to, and eventually becomes larger than, the equivalent one-loop uncertainty; this is due to the additional electroweak-scale logs introduced at the two-loop level, and their tendency to blow up as $\hat{M}$ drops significantly below $m_t$.

The light green uncertainty band is the naive estimate \cite{Ibe2013} of the missing two-loop contribution expected from loops involving the top quark:
\begin{align}
\frac{\alpha_{\text{EM}}^2 \,m_t}{16\pi \sin^4(\theta_W)} \sim 4 \ \text{MeV}. \label{eqn:error_estimate}
\end{align}
As we can see in Figure \ref{fig:wino_two_loop}, this estimate does indeed give a reasonable rough estimate of the uncertainty on the one-loop result.

\begin{table*}[tp]
\centering
\caption{The effect of uncertainties in input parameters on the mass splitting and decay lifetime in the wino and MDM models.  The effect on the decay lifetime is taken to be the difference between the upper and lower lifetimes normalised by the mean of the upper and lower values, expressed as a percentage.}\label{tab:uncertainty_analysis}
\footnotesize{
\vspace{0.4cm}
\begin{tabular}{cll@{\hspace{4mm}}cll}
 & \multicolumn{2}{c}{Wino model ($\hat{M}=3\,$TeV)} & & \multicolumn{2}{c}{MDM ($\hat{M}=9.6\,$TeV)}  \\
 \cline{2-3}  \cline{5-6}
& Change in & Change in & & Change in & Change in  \\
Parameter & $\Delta M$ (MeV) & lifetime (\%) & & $\Delta M^+$ ($\Delta M^{++})$ (MeV) & lifetime (\%) \\
\hline
$1/\alpha^{\overline{MS}}(m_Z)$   											&$0.0919$     & $0.310$ && $0.101$ ( $0.402$ )  & $0.348$ ( $0.209$ ) \\
 \phantom{$^{\overline{MS}}$}$m_t$\phantom{$^{\overline{MS}}$}  		& $0.192$       &$0.647$ &&  $0.175$ ( $0.699$ ) & $0.604$ ( $0.364$ )  \\
\phantom{$^{\overline{MS}}$}$m_h$\phantom{$^{\overline{MS}}$}  		& $0.0124$     &$ 0.0417$ &&  $0.017$ ( $0.068$ )  &  $0.0588$ ( $0.0354$ )   \\
\phantom{$^{\overline{MS}}$}$m_W$\phantom{$^{\overline{MS}}$}  		& $\num{8.22e-8}$  & $\num{2.77e-7}$&&  $\num{4.6e-9}$ ( $\num{1.85e-8}$ )   &  $\num{1.59e-8}$ ( $\num{9.65e-9}$ )  \\
 \phantom{$^{\overline{MS}}$}$m_Z$\phantom{$^{\overline{MS}}$}  		& $0.00936$   &$0.0316$&& $0.00467$ ( $0.0187$ )  & $0.0162$ ( $0.00974$ )  \\
 \phantom{$^{\overline{MS}}$}$m_e$\phantom{$^{\overline{MS}}$}  		& $\num{8.23e-6}$ & $\num{2.78e-5}$&&  $\num{2.04e-5}$ ( $\num{8.15e-5}$ )   & $\num{7.05e-5}$ ( $\num{4.24e-5}$ )   \\
\phantom{$^{\overline{MS}}$}$m_{\mu}$\phantom{$^{\overline{MS}}$}  	& $\num{3.69e-9}$ & $\num{1.25e-8}$&& $\num{9.87e-9}$ ( $\num{3.95e-8}$ ) & $\num{3.41e-8}$ ( $\num{2.06e-8}$ )   \\
\phantom{$^{\overline{MS}}$}$m_{\tau}$\phantom{$^{\overline{MS}}$}  	&  $\num{3.551e-6}$ & $\num{1.199e-5}$&&  $\num{3.37e-06}$ ( $\num{1.35e-5}$ )  & $\num{1.16e-05}$ ( $\num{7.01e-6}$ )  \\
$m_d^{\overline{MS}}(2\,\text{GeV})$  										& $\num{1.85e-4}$  &$ 0.000623 $&& $0.000845$ ( $0.00338$ )  & $0.00292$ ( $0.00176$ ) \\
$m_u^{\overline{MS}}(2\,\text{GeV})$     										& $\num{3.0927e-4}$    &$ 0.00104$&&   $0.00477$ ( $0.0191$ )  & $0.0165$ ( $0.00994$ ) \\
$m_s^{\overline{MS}}(2\,\text{GeV})$  										& $\num{8.467e-5}$   &$0.000286$&&  $0.001$ ( $0.00402$ ) & $0.00348$ ( $0.00209$ )   \\
$m_c^{\overline{MS}}(m_c)$ 													& $0.00176$    &$0.00595$&& $0.0017$ ( $0.00679$ )  & $0.00587$ ( $0.00354$ ) \\
$m_b^{\overline{MS}}(m_b)$  											     &  $0.0007539$ & $0.00255$&&  $0.00195$ ( $0.0078$ )  &  $0.00674$ ( $0.00406$ )  \\
$\alpha_S^{\overline{MS}}(m_Z)$  										     &  $0.00224$              &$0.00759$&& $0.00436$ ( $0.0174$ )   & $0.0151$ ( $0.00908$ )   \\
$Q$    																		     & $0.304$ 	    & $1.03$&& $0.242$ ( $0.969$ )   & $0.839$ ( $0.505$ )   \\
\hline\end{tabular}
}
\end{table*}

In Table \ref{tab:uncertainty_analysis} we present a detailed analysis of the uncertainties entering into this calculation.  As there are several uncorrelated uncertainties to include, we simply consider the effect of each individually.  The effect of including light quark masses is a $+0.0532\,$MeV change in the mass splitting, resulting in a $0.180$\% decrease in the lifetime.  The parameter with the largest effect on the mass splitting is the renormalisation scale.  Although this uncertainty is greatly reduced at the two-loop level (as seen in Figure \ref{fig:wino_two_loop}), it is still the dominant contribution.  We find that the uncertainties on the top mass and electromagnetic coupling also induce an $\mathcal{O}(0.1)$\,MeV uncertainty in the mass splitting.  All other parameters have negligible impacts on the mass splitting.  Although including the light quarks does slightly increase $\Delta M$, the uncertainties on these masses have almost no impact on the result.  Finally, we note that the strong coupling even has some influence, which is entirely indirect through the calculation of the spectrum, as this coupling is not directly involved in the wino mass calculation.

In the right panel of Figure \ref{fig:wino_two_loop} we present the decay lifetime of the charged component in units of mm/$c$, as a function of the degenerate mass $\hat{M}$.  The charged component decays as $\chi^+\rightarrow X\chi^0$, which is dominated by channels where $X$ is either a pion, an electron+neutrino or muon+neutrino pair.

The decay width for the pion channel in an electroweak multiplet with total weak isospin $j$, with eigenstates $\chi_{I}$ where $I\in\{-j,-j+1,.\,.\,.\,,j-1,j\}$, is given by
\cite{2015JHEP...07..074D}
\begin{align}
\Gamma\left(\chi_{I+1}\rightarrow\chi_{I}\pi^+\right) = T^2_+\frac{G_F^2\Delta M^3 V_{\text{ud}}^2f_{\pi}^3}{\pi}\sqrt{1-\frac{m_{\pi}^2}{\Delta M^2}}, \label{eqn:pi_width}
\end{align}
where $T_+=\sqrt{j(j+1)-I(I+1)}$, $f_\pi=130.2\pm1.7$\,MeV, $|V_{\text{ud}}|= 0.97417\pm0.00021$ \cite{Patrignani2016,2015arXiv150902220R} and $m_{\pi}$ is the pion mass. $T_+^2$ is equivalent to $(n^2-1)/4$ for $I=0$, for a representation of dimension $n$, as given in Ref.~\cite{Cirelli2009}, however for the MDM case we will need this more general expression.  For wino dark matter we have $j=1$ and $I=0$ to give $\Gamma(\chi^+\rightarrow\chi^0\pi^+)$.

For $\Delta M\approx 170\,\mathrm{MeV} > m_{\pi}$ the pion decay is the dominant channel, with a 97.7\% branching fraction \cite{Cirelli2009}.  The other kinematically-allowed channels are the electron-neutrino and muon-neutrino ones, which have widths
\begin{align}
\Gamma^{\chi^+}_{e}=T^2_+\frac{G_F^2\Delta M^5}{15\pi^2} \label{eqn:e_width}
\end{align}
and $\Gamma^{\chi^+}_{\mu}=0.12\Gamma^{\chi^+}_e$.  The expected lifetime of the charged component is thus $\tau=(\Gamma^{\chi^+}_e+\Gamma^{\chi^+}_{\mu}+\Gamma^{\chi^+}_{\pi})^{-1}$.  The large step in the decay lifetime in Figure \ref{fig:wino_two_loop} is where $\Delta M>m_{\pi}$ and the pion channel opens, and the smaller step is due to the muon channel opening.  These can be seen clearly as branching fractions in Figure \ref{fig:BFs}.

The most phenomenologically interesting mass range for pure wino-like neutralino dark matter is $\hat{M}\sim 3\,$TeV, as this would give the correct dark matter relic abundance \cite{Hisano2007,Hryczuk2011}.  For this value and assuming $Q=m_t$, the two-loop mass splitting is $164.5\,$MeV, compared to the one-loop value of $167.5\,$MeV.  This difference in mass splitting represents a $9.7\,$\% increase in the decay lifetime of the chargino when going from the one-loop to the two-loop calculation.  For other masses, this ratio can be larger, depending on the dominant decay channel.  For example, for a wino of $70\,$GeV mass, the one and two-loop mass splittings are $142.3\,$MeV and $145.5\,$MeV respectively, with a increase in the lifetime of 40.1\%.  Thus, although the difference in the mass splittings is approximately the same ($\sim$3\,MeV), as we can see in Figure \ref{fig:BFs}, this mass value is exactly where the pion channel opens up, so the effect on the lifetime in this range is far more significant.

We now offer an updated fit, using the latest values in Table \ref{table:params}, two-loop RGEs and non-zero light quark masses,
\begin{widetext}
\begin{eqnarray}
\frac{\Delta M}{1~\MEV} &=&  - 413.7
+305.7 \left(\ln \frac{\Mp^0}{1~\GEV}\right)
-60.96 \left(\ln\frac{\Mp^0}{1~\GEV}\right)^2
 +5.429 \left(\ln\frac{\Mp^0}{1~\GEV}\right)^3\label{eqn:fit1a}
-0.182 \left(\ln\frac{\Mp^0}{1~\GEV}\right)^4.
\end{eqnarray}
\end{widetext}
This fit is valid over the range $100\,\mathrm{GeV} \le M^0_\mathrm{pole} \le 4$\,TeV.  The effect of including light quark masses is a small positive shift in $\Delta M$, and 2-loop RGEs a smaller negative shift, with a total difference of approximately $-0.03\%$.

\subsection{The MDM quintuplet}

The MDM quintuplet has two mass splittings.  The first, $\Delta M^{+}\equiv \Mp^{+}-\Mp^0$, is analogous to $\Delta M$ in the wino model, with a one-loop value of $\mathcal{O}(170)\,$MeV.  The second, $\Delta M^{++}\equiv \Mp^{++}-\Mp^0$, between the neutral and doubly-charged component, has a value of $\mathcal{O}(670)\,$MeV at one loop.  In this section we present the first analysis of these mass splittings at the two-loop level and the subsequent decay lifetimes of the charged components.  In Section \ref{sec:comparison} we discuss the differences between the charged/neutral component mass splitting in the MDM and wino models.

\begin{figure*}
\centering
\includegraphics[width=0.5\textwidth]{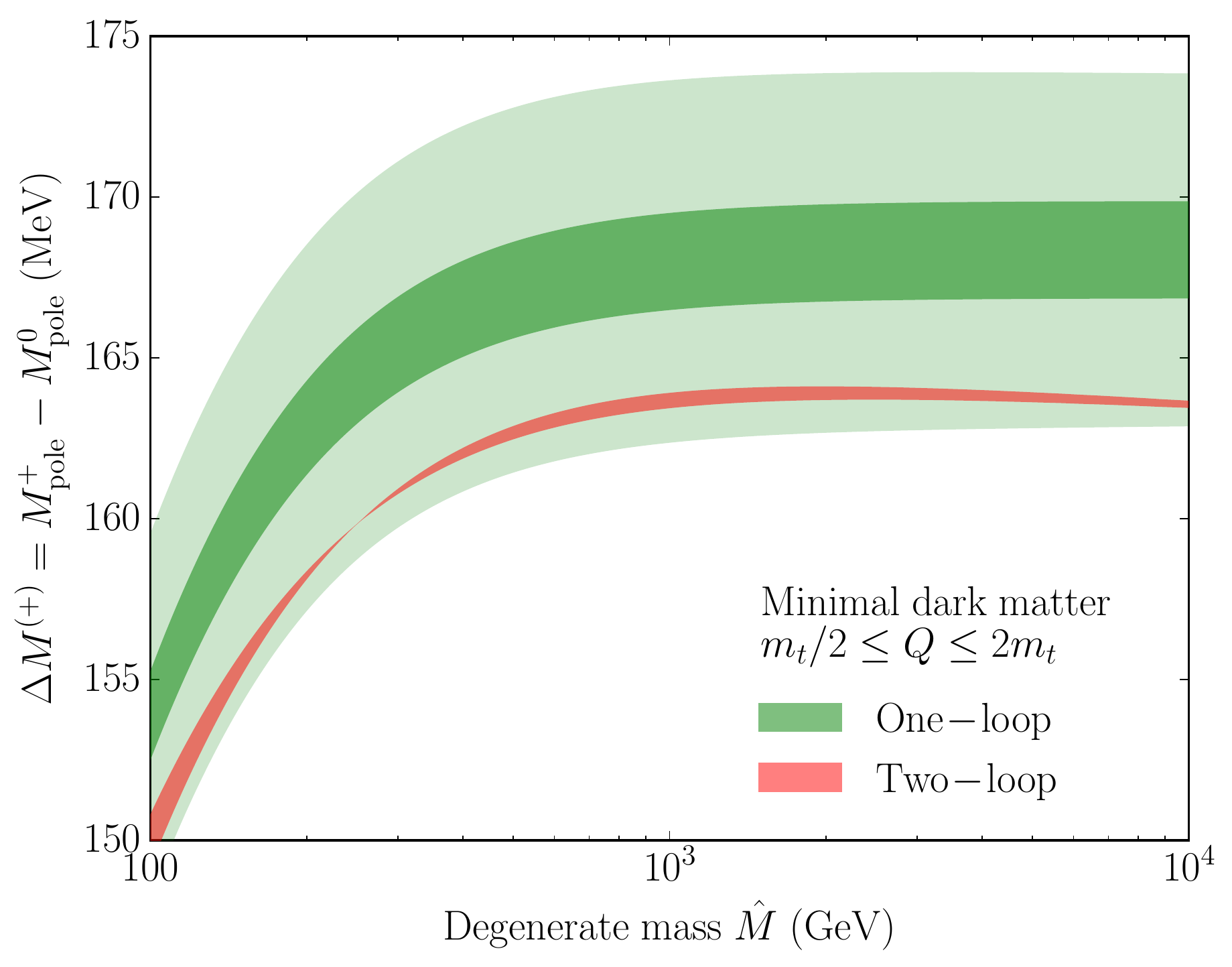}\includegraphics[width=0.5\textwidth]{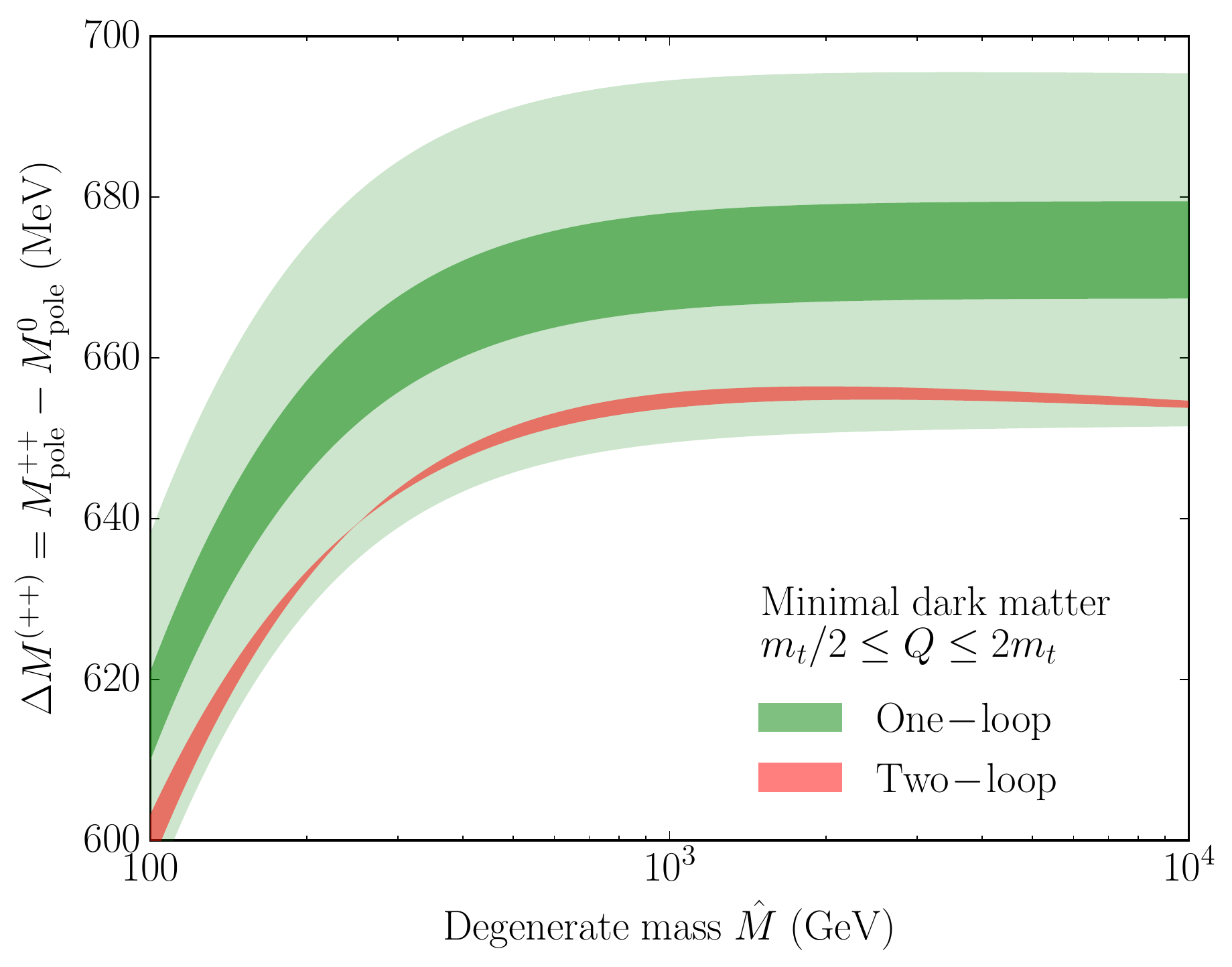}
\caption{The two-loop mass splittings between the charged and neutral components (\textit{left}) and the doubly-charged and neutral components (\textit{right}) in the MDM model as a function of the degenerate tree-level \MSbar mass.  The dark green and red bands are the range of the one and two-loop mass splittings respectively when $Q$ is varied continuously between $m_t/2$ and $2m_t$.  The light green band is the estimated uncertainty on the one-loop result using Eq.~(\ref{eqn:error_estimate}).}\label{fig:deltam_mdm}
\end{figure*}

In Figure \ref{fig:deltam_mdm} we present the two-loop mass splittings between the neutral and charged (\textit{left panel}) and the neutral and doubly-charged (\textit{right panel}) components.  The dominant uncertainty, resulting from the choice of renormalisation scale, is indicated by the dark shaded regions at one loop (\textit{dark green}) and two loops (\textit{red}), where $Q$ has been varied continuously between $m_t/2$ and $2m_t$.  Once again we see a significant reduction in the uncertainty at the two-loop level, at least for moderate and large multiplet masses; at lower multiplet masses ($\hat{M}\lesssim100$\,GeV), the two-loop uncertainty grows due to the additional electroweak-scale logs introduced at the two-loop level, just as in the triplet case.  The light-green band is the naive estimate of the missing two-loop contribution, where we use Eq.\ (\ref{eqn:error_estimate}) for $\Delta M^+$, and multiply this by a factor of four for $\Delta M^{++}$, based on the generic charge-dependent pre-factors for one-loop electroweak mass splitting in Eq.\ (4) of Ref.~\cite{Yamada2010}.

In Table \ref{tab:uncertainty_analysis}, we also give a detailed presentation of the uncertainties entering into the two-loop calculation in the MDM model.  Again, we consider the effect of each uncertainty individually.  As in the wino case, the parameter with the largest effect on the mass splitting is the renormalisation scale, but its effect is greatly reduced by going to two loops (Figure \ref{fig:deltam_mdm}).  The top mass and electromagnetic coupling are again responsible for an $\mathcal{O}(0.1)$ MeV uncertainty in the mass splittings.  All other parameters have negligible impacts on the splittings.  Including the masses of light quarks results in a $+0.0125\,$MeV change in $\Delta M^+$ and a $0.0499\,$MeV change in $\Delta M^{++}$, which translate into $0.0432$\% and $0.0258$\% reductions in the respective lifetimes of the charged and doubly-charged states.  As with the triplet, although finite light quark masses affect $\Delta M$, the uncertainties on those masses have little impact -- and the strong coupling has some influence via the calculation of the spectrum (on the order of $0.01$\%).

\begin{figure*}
\centering
\includegraphics[width=0.7\textwidth]{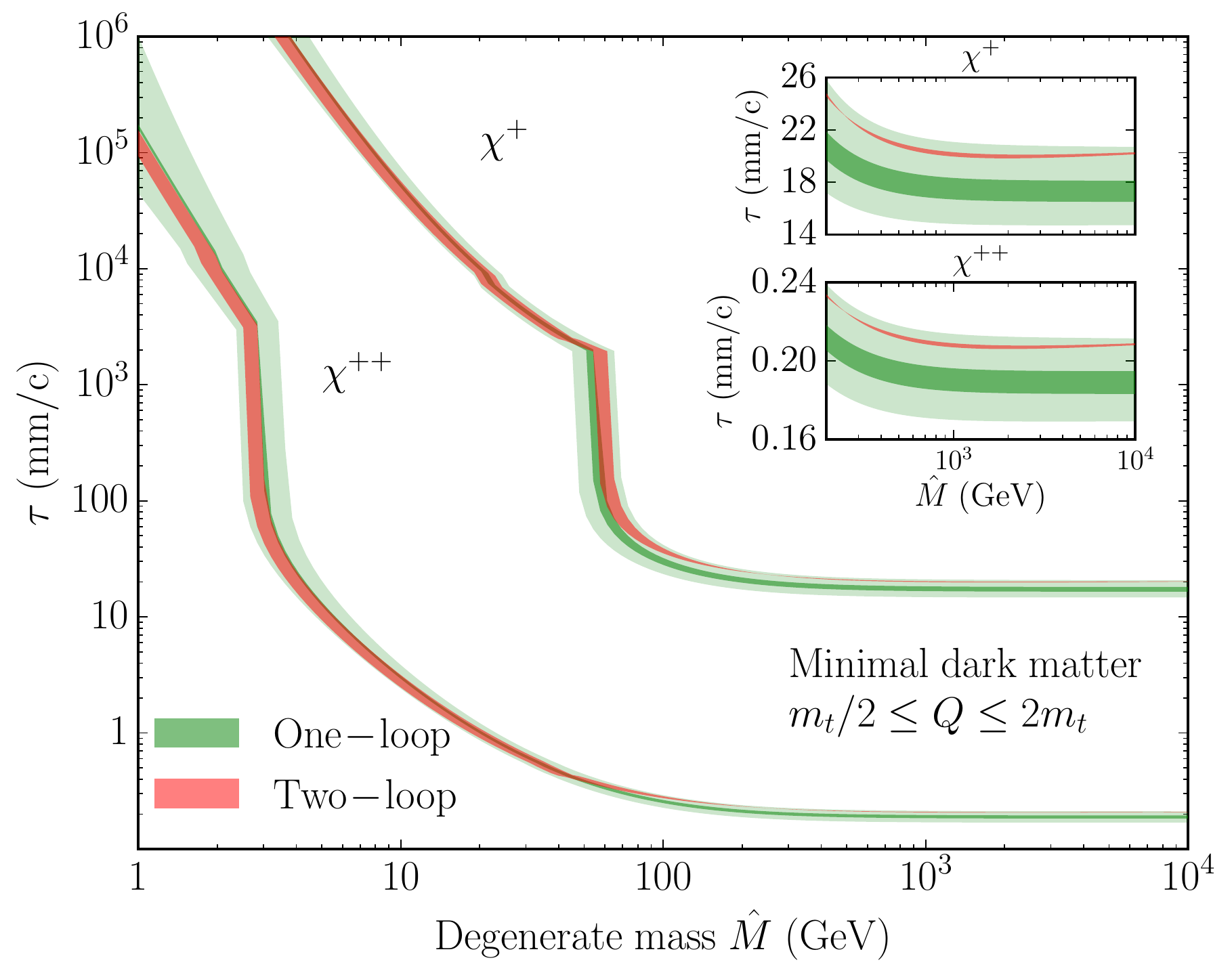}
\caption{The decay lifetimes of the charged and doubly-charged components in the MDM model as a function of the degenerate tree-level \MSbar mass.  The dark green and red bands are the respective ranges of the one and two-loop mass splittings when $Q$ is varied continuously between $m_t/2$ and $2m_t$.  The light green band is the estimated uncertainty on the one-loop result using Eq.~(\ref{eqn:error_estimate}).}\label{fig:decays_mdm}
\end{figure*}

In Figure \ref{fig:decays_mdm} we plot the decay lifetimes of the charged and doubly-charged components.  The lifetime of the charged component can be computed using Eqs.~(\ref{eqn:e_width}) and (\ref{eqn:pi_width}) with $j=2$ and $I=0$; the calculation is the same for the doubly-charged component, but with $I=1$ instead.  The doubly-charged component has an additional decay channel via the process $\chi^{++}\rightarrow\chi^+K^+$, where $K^+$ is a kaon.   We take the partial decay width to the kaon channel to be
\begin{align}
\Gamma_{K^+} = T^2_+\frac{G_F^2\Delta M^3 V_{\text{us}}^2f_{K^+}^3}{\pi}\sqrt{1-\frac{m_{K^+}^2}{\Delta M^2}}, \label{eqn:k_width}
\end{align}
where $f_{K^+}=155.6\pm0.4$\,MeV, $|V_{\text{us}}|=  0.2248\pm0.0006$ \cite{Patrignani2016,2015arXiv150902220R} and $m_{K^+}$ is the kaon mass.

\begin{figure*}
\centering
\includegraphics[width=0.5\textwidth]{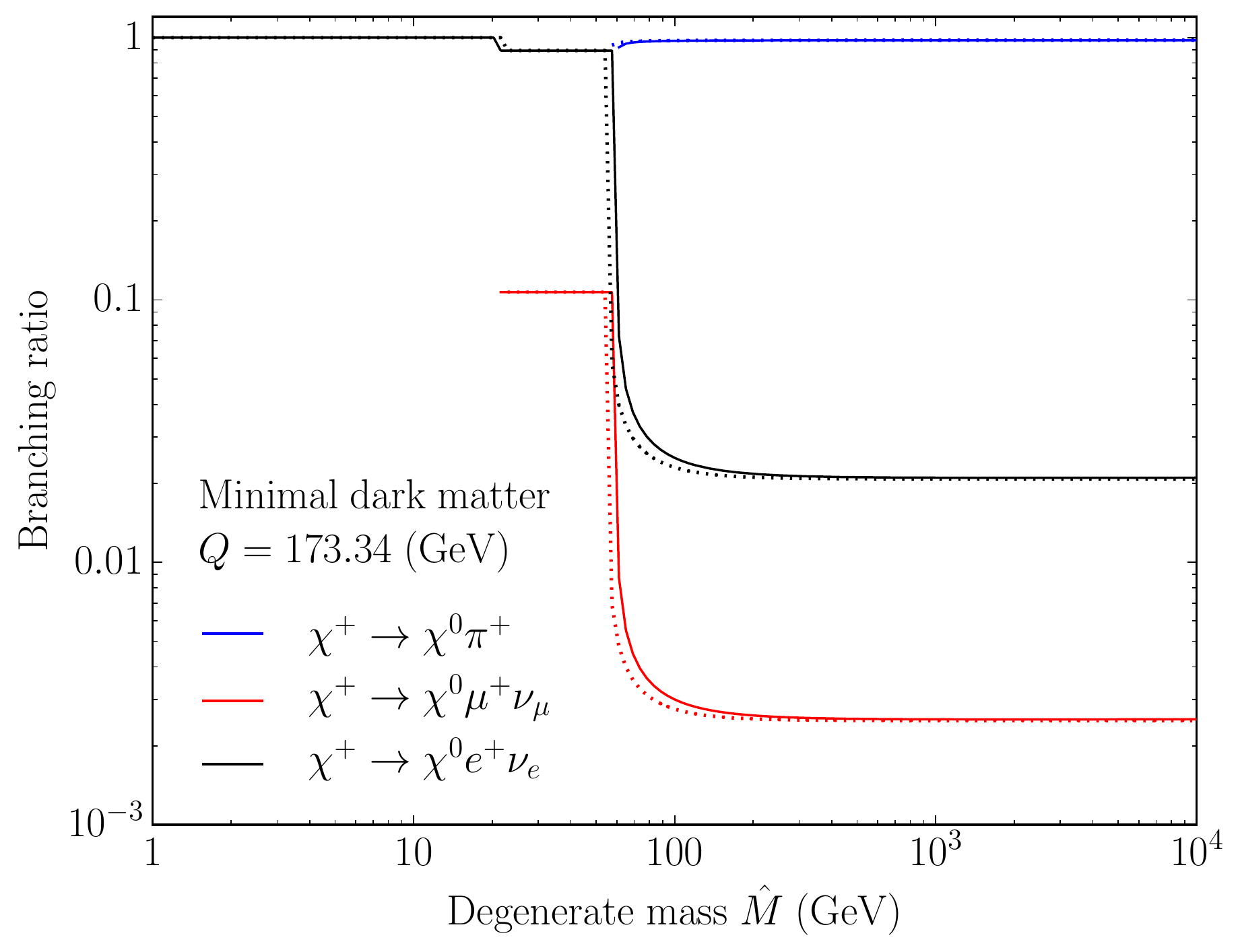}\includegraphics[width=0.5\textwidth]{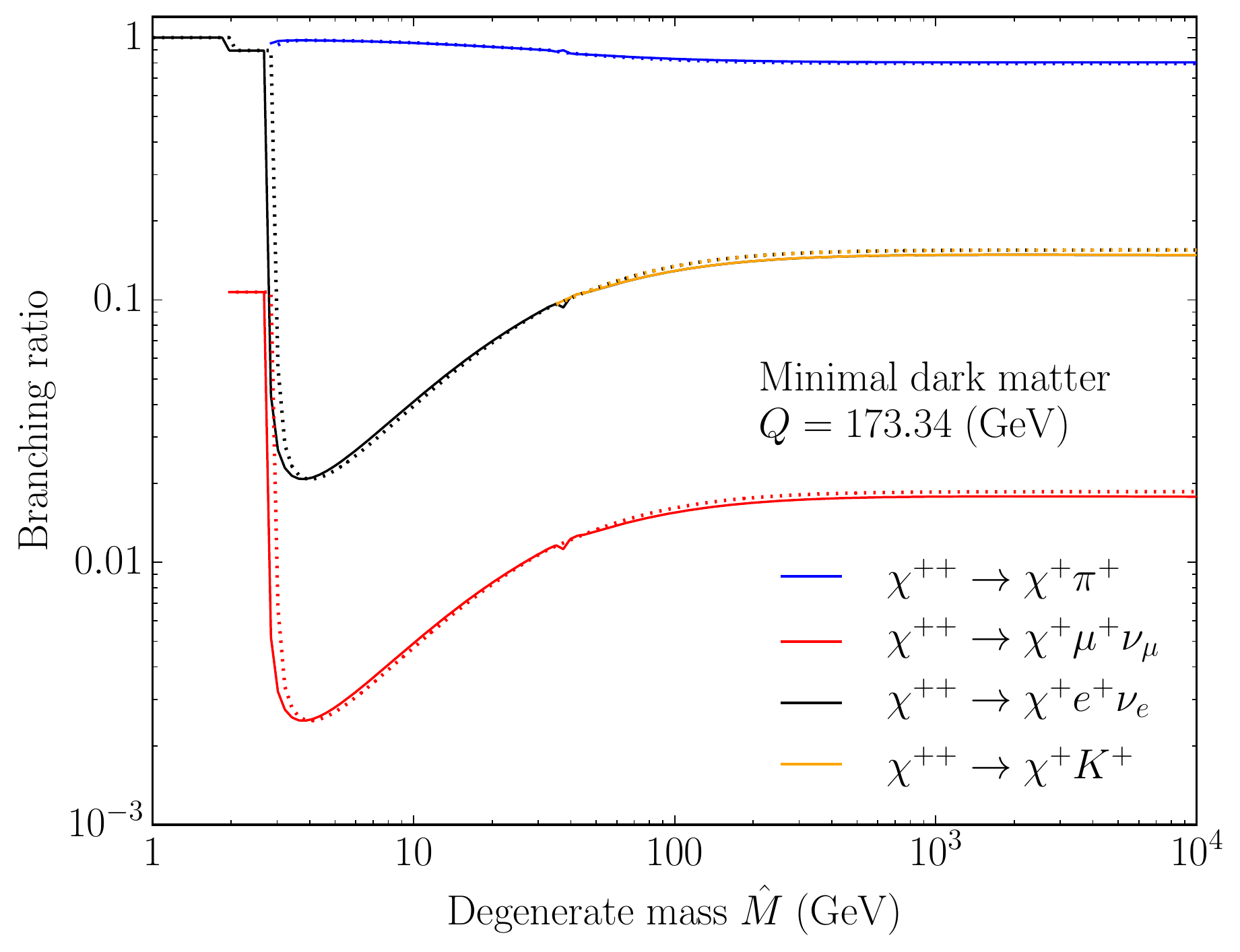}
\caption{Branching fractions in the MDM model for the $\chi^+\rightarrow X \chi^0$ (\textit{left}) and $\chi^{++}\rightarrow X \chi^+$ (\textit{right}) processes, where $X \in \{e^+\nu_e,\mu^+\nu_\mu,\pi^+,K^+\}$.  The solid lines are the branching fractions using the two-loop mass splitting and dotted lines are the results using the one-loop result, both evaluated at $Q=m_t=173.34\,$GeV.}\label{fig:BR_mdm}
\end{figure*}

The most phenomenologically interesting mass range for MDM is $\hat{M}\sim 9.6\,$TeV, as this would give the correct dark matter relic abundance \cite{Cirelli2007}.  For this value the two-loop mass splittings are $\Delta M^+ = 163.6\,$MeV and $\Delta M^{++} =  654.3\,$MeV, which can be compared with the one-loop values of $168.3\,$MeV and  $673.4\,$MeV respectively, for a choice of $Q=m_t$.  This difference in mass splitting represents a 15.5\% change in the decay lifetime of the charged component when going from the one-loop to the two-loop calculation, and a 9.78\% change in the decay lifetime of the doubly-charged component.  Like in the wino model, this ratio will be larger at different mass values, depending on the dominant decay channel.  One important new feature in this calculation is the opening up of the kaon channel, which we indicate with the orange line in the right panel of Figure \ref{fig:BR_mdm}.

These two-loop mass splitting results can be reproduced using the following fitting formulae.  For $\Delta M^{(+)}$, we have
\begin{widetext}
\begin{eqnarray}
\frac{\Delta M^{(+)}}{1~\MEV} &=&  - 328.6
+250.1 \left(\ln \frac{\Mp^0}{1~\GEV}\right)
- 47.7 \left(\ln\frac{\Mp^0}{1~\GEV}\right)^2
 +4.049 \left(\ln\frac{\Mp^0}{1~\GEV}\right)^3
-0.1292 \left(\ln\frac{\Mp^0}{1~\GEV}\right)^4. \label{eqn:fit2}
\end{eqnarray}
and for $\Delta M^{(++)}$, we have
\begin{eqnarray}
\frac{\Delta M^{(++)}}{1~\MEV} &=&  - 1314
+1000 \left(\ln \frac{\Mp^0}{1~\GEV}\right)
- 190.7 \left(\ln\frac{\Mp^0}{1~\GEV}\right)^2
+16.18 \left(\ln\frac{\Mp^0}{1~\GEV}\right)^3
-0.5162 \left(\ln\frac{\Mp^0}{1~\GEV}\right)^4. \label{eqn:fit3}
\end{eqnarray}
\end{widetext}
These formulae are valid for values of $M^0_\mathrm{pole}$ between $100\,$GeV and $10\,$TeV.

\subsection{Differences between triplet and quintuplet models}\label{sec:comparison}

The two-loop loop mass splitting between the charged and neutral multiplet component is not identical in the triplet and quintuplet models.  At the one-loop level this mass splitting is the same in both representations, yet when we go to the next loop order there are subtle differences.  In this section we discuss these differences and determine which diagrams are responsible.

In the MDM model, for multiplet masses $\gtrsim 1$\,TeV we see a decrease in the two-loop mass splitting.  In the two-loop wino result, and in the one-loop case for both models, we see a constant mass splitting in the limit of large $\hat{M}$.  In the one-loop case, this can be seen directly from the difference of the one-loop self energies (given in Appendix \ref{sec:self_energies}), and the fact that we do not apply threshold corrections (as they are technically of higher loop order).  If we were to include threshold corrections to the one-loop result, we would see a similar decrease in the mass splitting for large $\hat{M}$, as we would be introducing extra logarithmic terms with nothing to cancel them.

In the wino model the constant mass splitting at large $\hat{M}$ is the result of a cancellation between these threshold corrections and one specific set of diagrams.  These are specifically the corrections to the gauge boson propagators coming from the new multiplet fermions.  The diagrams that contribute to the gauge boson propagators are are all those in Figure \ref{fig:Feynman_diagrams2} and the first counter-term diagram in Figure \ref{fig:Feynman_diagramsCT}.  Ref.\ \cite{Yamada2010} asserts that this cancellation occurs exactly for \textit{all} SU(2) multiplets, and therefore goes on to ignore threshold corrections and the influence of the multiplet fermions on the gauge boson propagator.  Our calculations show that this cancellation does indeed occur for the triplet, but that the resulting logs do \textit{not} perfectly cancel in the quintuplet case.  The fact that the mass splitting is almost flat in the large $\hat{M}$ limit indicates that \textit{most} of the logs have cancelled (as e.g.\ neglecting threshold corrections results in a clear logarithmic increase in the splitting with increasing $\hat{M}$) -- but some small residual term of the form $-\log(\hat{M}/Q)$ remains.

\begin{figure*}
\centering
\includegraphics[width=0.5\textwidth]{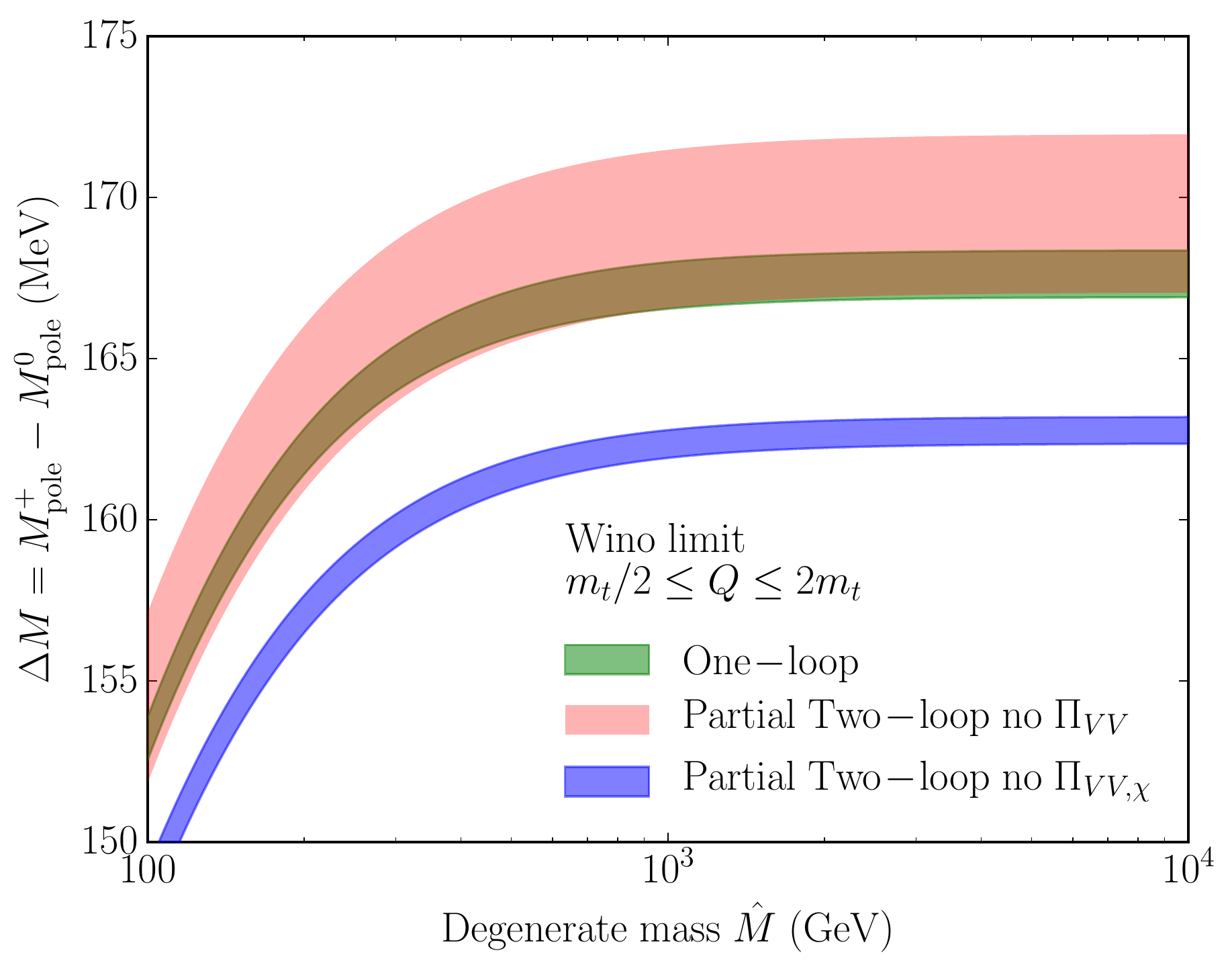}\includegraphics[width=0.5\textwidth]{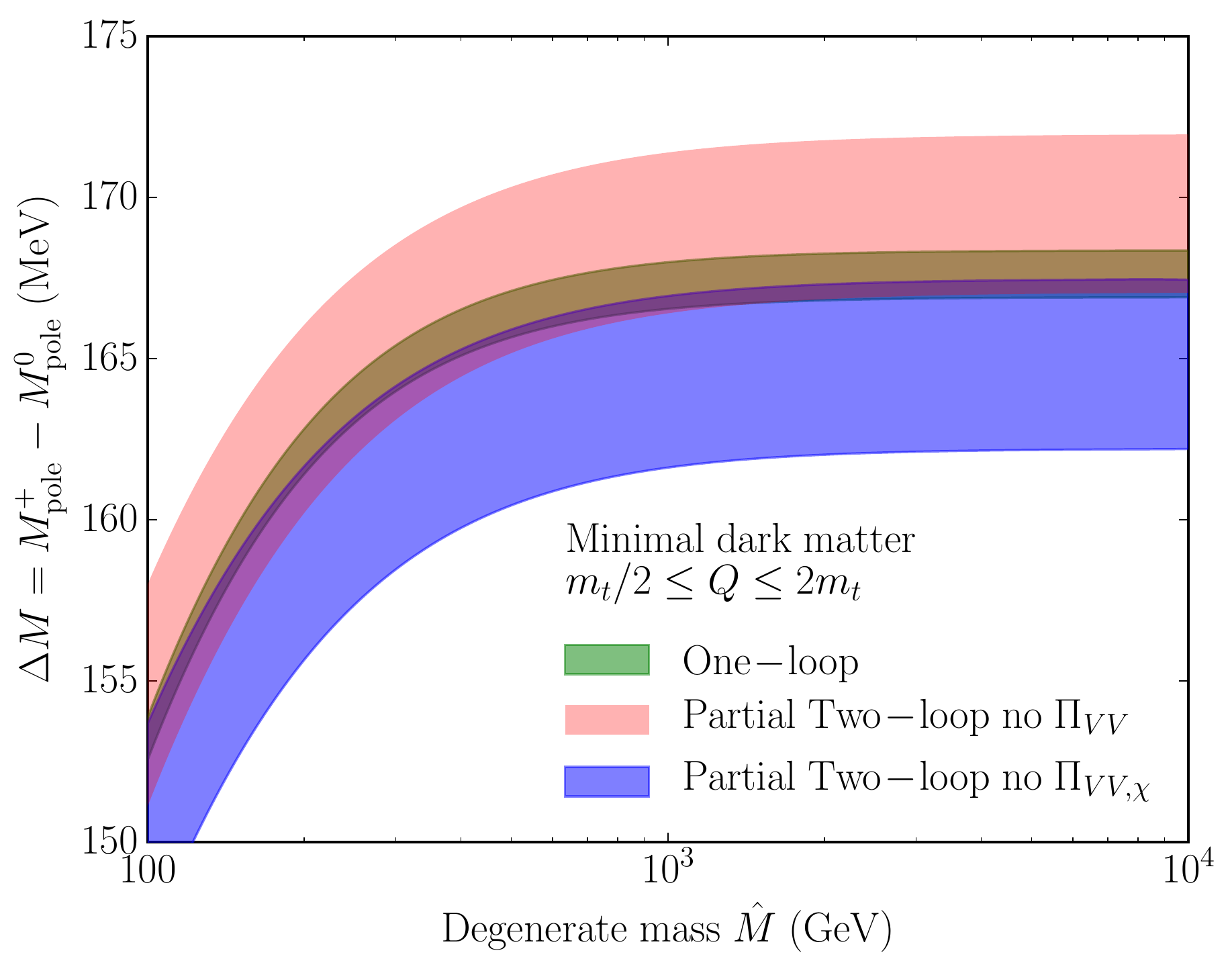}
\caption{
The one-loop (green band), partial two-loop (red band) and extended partial two-loop (blue band) mass splittings in the wino limit of the MSSM (\textit{left}) and the MDM quintuplet model (\textit{right}). The partial two-loop mass splitting is computed using self energies constructed from diagrams in Figure \ref{fig:Feynman_diagrams} and all except the left-most diagram in Figure \ref{fig:Feynman_diagramsCT}, that is, all two-loop diagrams except those that include a correction to a gauge boson propagator.  The extended partial two-loop mass splitting is calculated with all two-loop diagrams except those that include a correction to a gauge boson propagator by $\chi$ fermions.  The coloured bands are determined by varying $Q$ continuously between $m_t/2$ and $2m_t$.  For these calculations, we neglect threshold corrections and all running of parameters.}\label{fig:multiplet_compare}
\end{figure*}

To illustrate this point, we can construct a partial two-loop mass-splitting calculation with the terms responsible for the residual logs excluded.  First, we construct two-loop amplitudes by neglecting threshold corrections and excluding all contributions to the gauge boson self-energy, i.e. all diagrams in Figure \ref{fig:Feynman_diagrams2} and the first in Figure \ref{fig:Feynman_diagramsCT}.  In Figure \ref{fig:multiplet_compare}, we plot the resulting partial two-loop mass splittings in each model as `Partial Two-loop no $\Pi_{VV}$', along with the one-loop results.  We see that the results are indeed identical at large $\hat{M}$.  We can also see that this incomplete subset of diagrams misses some important cancellations of scale-dependent logarithmic terms, as the uncertainty from scale dependence in the resulting two-loop splitting is much larger in the partial two-loop amplitude compared to the full two-loop result.

To investigate further, we next exclude \textit{only} those diagrams where the multiplet fermions contribute to the gauge boson propagator, i.e.\ the versions of the second diagram in Figure \ref{fig:Feynman_diagrams2} with a $\chi$ fermion in the upper loop.  Continuing to neglect threshold corrections, we then recompute the corresponding counter-term (the first in Figure \ref{fig:Feynman_diagramsCT}) with the same contributions removed from the gauge boson propagator, and recompute the mass splitting.  We refer to this extended partial amplitude as `Partial Two-loop no $\Pi_{VV,\chi}$' in Figure \ref{fig:multiplet_compare}.  The splitting is still flat at large $\hat{M}$ in both models, albeit with a larger scale dependence in the MDM model due to the contributions of a large number of additional diagrams (relative to the wino model) in Fig.\ \ref{fig:Feynman_diagrams2} with $\chi^{\pm\pm}$ in the lower internal propagator.  The flatness of the extended partial two-loop result at large $\hat{M}$ shows that the uncanceled logarithms in the full quintuplet calculation specifically arise from the failure of the threshold corrections to fully cancel the logs from the contribution of the multiplet fermions to the gauge boson self-energies.  Unlike light quark masses, which increase the mass splitting (Figure \ref{fig:Ibe_compare}), the addition of multiplet fermions reduces the splitting, as the two types of fermions enter into the gauge boson self energies with opposite signs.  The fact that the mass splitting ultimately turns down in the MDM quintuplet therefore indicates that the impacts of the multiplet fermions on the gauge boson self-energies dominate over the threshold corrections in this model.

That the logarithms do not fully cancel in the quintuplet model suggests that they will also not completely cancel for higher-dimensional representations of SU(2).

\section{Conclusion}
\label{sec5}

We have presented a two-loop calculation of mass splitting in electroweak multiplets, in the wino limit of the MSSM and in the MDM fermionic quintuplet model.  In the wino model, we showed that our calculation is in agreement with the previous two-loop calculation.  We improved on the previous calculation by using two-loop RGEs and including finite masses for light quarks.

We also presented the first complete two-loop calculation of the splitting in the MDM quintuplet model, showing that it is not constant in the limit of large multiplet masses.  This is contrary to the triplet case, and the naive expectation from the one-loop result.  This result comes from the influence of the additional heavy fermions on the gauge boson self energies, and subsequently the two-loop self energies of the multiplet.  As the mass of the multiplet increases, so does its effect on the mass splitting through these diagrams.

The two-loop corrections that we present here are phenomenologically relevant, resulting in a $\sim$10\% change in the lifetime of the charged components in both models.  This is in agreement with previous calculations for wino dark matter \cite{Ibe2013}.  It is similarly important to include the two-loop radiative corrections presented here when considering disappearing track searches for MDM.

\begin{acknowledgments}
We would like to thank Ryosuke Sato and Sanjay Bloor for useful discussions, Vladyslav Shtabovenko and Stephen Martin for helpful correspondence regarding technical aspects of two-loop calculations and the use of \feyncalc and \tsils, respectively, and Peter Athron and Alex Voigt for helpful correspondence about \flexiblesusys, \sarah and threshold corrections.  We thank Julian Heeck for helpful comments on the submitted version of this paper. JM is supported by the Imperial College London President's PhD Scholarship and PS by STFC (ST/K00414X/1, ST/N000838/1, ST/P000762/1).
\end{acknowledgments}

\begin{widetext}
\appendix

\section{One-loop self energies and counter-term couplings}\label{sec:self_energies}

Here we present the one-loop self energies and counter-term couplings required for the computation of the two-loop mass splitting. The two-loop multiplet self energies are omitted, but a \CC computer code with the self energies expressed in the form described in Sec.~\ref{sec:calc_method}, as coefficients of basis integrals, is available on request and will be made public as part of a future code release.

One-loop self energies for the multiplet components are presented in Sections \ref{app:mssm_n} and \ref{app:mssm_c} for the wino limit of the MSSM, and Sections \ref{app:mdm_n}, \ref{app:mdm_c} and \ref{app:mdm_cc} for the MDM model.  Counter-term couplings for the new two and three-point vertices are provided in Sections \ref{app:mssm_ct} and \ref{app:mdm_ct} for the wino and MDM models, respectively.

To compute the two-loop amplitudes in the left-most diagram of Figure \ref{fig:Feynman_diagramsCT}, we need to determine the counter-term couplings for the gauge boson propagators.  This is achieved by computing the one-loop gauge boson self energies and setting the counter-term couplings such that the UV divergences cancel.  In both the wino limit of the MSSM and MDM, the self energies of the electroweak gauge bosons are given by the SM contribution plus an additional one or two diagrams from the new multiplet.  Let the self energy of the gauge bosons be
\begin{align}
\Pi_{V_1V_2} = \Pi_{V_1V_2,SM} + \Pi_{V_1V_2,\chi\chi} + \delta_{Z,V_1V_2}(p^2-\hat{m}_V^2)-\delta_{M,V_1V_2}
\end{align}
where $V_i\in \{ W,Z,\gamma\}$, $\Pi_{VV,SM}$ is the SM contribution,  $\hat{m}_V$ is the boson mass when $V_1=V_2$ or zero otherwise and $\delta_{Z,V_1V_2}, \delta_{M,V_1V_2}$ are counter-term couplings.  The SM part, $\Pi_{VV,SM}$, which consists of the contributions from other gauge bosons, fermions, ghosts and Goldstone bosons can be found in multiple sources (see for example Refs.~\cite{Pierce1997, Yamada2010, Ibe2013}), so we do not reproduce them here.  The contributions to the gauge boson self energies from the new multiplet components are presented in Sections \ref{app:mssm_vv} and \ref{app:mdm_vv}, respectively, for the wino and MDM models.  We also provide the full counter-term couplings, including the SM contributions, for the gauge bosons in Sections \ref{app:mssm_ct} and \ref{app:mdm_ct}.

For one-loop self energies we need only two basis integrals, or Passarino-Veltman (PV) \cite{tHooft1979,Passarino1979} functions.  These integrals are defined as
\begin{eqnarray}
A(m) &=& 16\pi^2Q^{4-d}\int{d^dq\over i\,(2\pi)^d}{1\over
q^2+m^2+i\varepsilon}\\
B(p, m_1, m_2) &=&
16\pi^2Q^{4-d}\int{d^dq\over i\,(2\pi)^d}
{1\over\left[q^2+m^2_1+i\varepsilon\right]\left[
(q-p)^2+m_2^2+i\varepsilon\right]},
\label{B0 def}
\end{eqnarray}
where we use $d=4-2\epsilon$.  The complex solutions to these integrals can be expressed analytically; see Ref.~\cite{Pierce1997} for more details.

Throughout this appendix, the separation of fermion self energies into the form $\Sigma(p^2)=\Sigma_K(p^2)\slashed{p}+\Sigma_M(p^2)$ is manifest in the form of the coefficients.  All self energies are in the Feynman-'t Hooft ($\xi=1$) gauge and we define $\kappa\equiv 1/(16\pi^2)$.

\subsection{Wino model}

\subsubsection{Neutral component}\label{app:mssm_n}
The self energy of the neutral component $\chi^0$ is
\begin{align}
\begin{split}
\kappa^{-1}\Sigma^0(p^2) =& \, C^0_{A_\chi}\,A(\hat{M})  + C^0_{A_W}\,A(\hat{m}_W)  + C^0_{B_{\chi W}}\,B(\hat{M},\hat{m}_W) + C^0_0,
\end{split}
\end{align}
 with coefficients
 \begin{eqnarray}
&C^0_{A_\chi} &= -\frac{2g^2}{p^2} \slashed{p} \\
&C^0_{A_W} &= \frac{2g^2}{p^2} \slashed{p}\\
&C^0_{B_{\chi W}} &= \frac{2g^2}{\hat{m}_W^2p^2} \left( p^2+\hat{M}^2-\hat{m}_W^2   \right)\slashed{p} - 8g^2\hat{M}\\
&C^0_0 &= \left(-2g^2 + \delta_{\chi,Z}\right) \slashed{p}  + \left(4g^2+\delta_{\chi,M}\right)\hat{M}.
\end{eqnarray}

\subsubsection{Charged component}\label{app:mssm_c}

The self energy of the charged component $\mychi^+$ is given by
\begin{align}
\begin{split}
\kappa^{-1}\Sigma^+(p^2) = \, & C^+_{A_\chi}\,A(\hat{M})  + C^+_{A_W}\,A(\hat{m}_W)   + C^+_{A_Z}\,A(\hat{m}_Z)      \\ &  +  C^+_{B_{\chi\gamma}}\,B(\hat{M},0)     + C^+_{B_{\chi W}}\,B(\hat{M},\hat{m}_W) + C^+_{B_{\chi Z}}\,B(\hat{M},\hat{m}_Z)    + C^+_0,
\end{split}
\end{align}
with coefficients
\begin{eqnarray}
&C^+_{A_{\chi}}  & = -\frac{2g^2}{p^2} \slashed{p} \\
&C^+_{A_W}  & =  \frac{g^2}{p^2} \slashed{p} \\
&C^+_{A_Z}  & =  \frac{g^2\cos^2(\theta_W)}{p^2} \slashed{p}   \\
&C^+_{A_{\gamma}}  & =   \frac{g^2\sin^2(\theta_W)}{p^2} \slashed{p}  \\
&C^+_{B_{\chi W}}  & = \frac{g^2}{p^2}\left( p^2+\hat{M}^2-\hat{m}_W^2 \right) \slashed{p} - 4g^2\hat{M}  \\
&C^+_{B_{\chi\gamma}}  & =  \frac{\sin^2(\theta_W)g^2}{p^2}\left( p^2+\hat{M}^2\right) \slashed{p} - 4g^2\hat{M} \sin^2(\theta_W) \\
&C^+_{B_{\chi Z}}  & =   \frac{\cos^2(\theta_W)g^2}{p^2}\left( p^2+\hat{M}^2-\hat{m}_Z^2 \right) \slashed{p} - 4g^2\hat{M} \cos^2(\theta_W) \\
&C^+_0   & =  \left( -2g^2  +\delta_{\chi,Z}\right)\slashed{p} + \left(4g^2 + \delta_{\chi,M}\right)\hat{M}.
\end{eqnarray}

\subsubsection{Gauge bosons}\label{app:mssm_vv}

The multiplet contributions are given by
\begin{eqnarray}
&\Pi_{ZZ,\chi\chi} &= \frac{e^2\cot^2(\theta_W)}{36\pi^2}\Pi(\hat{M})\\
&\Pi_{\gamma\gamma,\chi\chi} &= \frac{e^2}{36\pi^2}\Pi(\hat{M})\\
&\Pi_{WW,\chi\chi} &= \frac{g^2}{36\pi^2}\Pi(\hat{M})\\ \label{eq:W_SE_chi}
&\Pi_{Z\gamma,\chi\chi} &= \frac{e^2\cot^2(\theta_W)}{36\pi^2}\Pi(\hat{M}),
\end{eqnarray}
where
\begin{align}
\label{eqn:pi}
\Pi(m) \equiv 3(p^2+2m^2)\, B(p,m,m) - p^2 - 6\, A(m) + 6m^2.
\end{align}

\subsubsection{Counter-term couplings}\label{app:mssm_ct}

The counter-terms $\delta_Z$ and $\delta_M$ required to cancel divergences arising from $B_0$ and $A_0$ are
\begin{eqnarray}
\delta_{\chi,Z}&=&4g^2\Delta\\
\delta_{\chi,M}&=&-16g^2\Delta,
\end{eqnarray}
where $\Delta \equiv 2/(4-d)-\gamma_\mathrm{E}+\log(4\pi)$ and $\gamma_\mathrm{E}$ is the Euler-Mascheroni constant.

Additional one-loop counter-terms are required to control divergences in the two-loop self energies. These are the counter-terms for the gauge-multiplet three-point vertices,
\begin{eqnarray}
\delta_{\cn\cn Z} &=& \frac{g^3}{4\pi^2}\Delta\\
\delta_{\cn\cp W} &=& \frac{\delta_{\cp\cp \gamma}}{\sin(\theta_W)} = \frac{\delta_{\cp\cp Z}}{\cos(\theta_W)} = -\frac{g^3}{2\pi^2}\Delta.
\end{eqnarray}

The gauge boson counter-term couplings are
\begin{eqnarray}
&\delta_{Z,WW}  &=  -\frac{13g^2}{96\pi^2}\Delta\\\
&\delta_{M,WW} &= \frac{g}{32\pi^2} \left[ -\sum_{i}c_im_i^2 + 13\hat{m}_W^2-6\hat{m}_Z^2\cos(2\theta_W)\right]\Delta\\\
&\delta_{Z,ZZ}  &=  \frac{g^2}{96\pi^2}\left[54\sin^2(\theta_W)-41\sec^2\theta_W+28\right]\Delta\\\
&\delta_{M,ZZ} &= -\frac{g\sec^2(\theta_W)}{96\pi^2} \left[  3\sum_i c_i m_i + \hat{m}_W^2\left(55-47\sec^2\theta_W-15\cos2\theta_W  \right)
  \right]\Delta\\\
&\delta_{Z,\gamma\gamma} & =  -\frac{9g^2\sin^2\theta_W}{16\pi^2}\Delta\\\
&\delta_{M,\gamma\gamma} & = 0\\
&\delta_{Z,Z\gamma} & =  \frac{g^2\tan\theta_W}{96\pi^2}\left(14-29\cos2\theta_W\right) \Delta\\\
&\delta_{M,Z\gamma} & =   -\frac{g^2\hat{m}_Z^2}{8\pi^2}\sin\theta_W\cos\theta_W\Delta\
\end{eqnarray}
where the summation is over all SM quarks and leptons, with
\begin{align}
m_i\in\{\hat{m}_u,\hat{m}_c,\hat{m}_t,\hat{m}_d,\hat{m}_s,\hat{m}_b,\hat{m}_{e},\hat{m}_{\mu},\hat{m}_{\tau}\} \label{eqn:mi}
\end{align}
and $c_i = 3$ for quarks and $1$ for leptons.

\subsection{Minimal dark matter}

\subsubsection{Neutral component}\label{app:mdm_n}
The self energy of the neutral component, $\mychi^0$, is
\begin{align}
\begin{split}
\kappa^{-1}\Sigma^0(p^2) =& \, C^0_{A_\chi}\,A(\hat{M})  + C^0_{A_W}\,A(\hat{m}_W)  + C^0_{B_{\chi W}}\,B(\hat{M},\hat{m}_W) + C^0_0
\end{split}
\end{align}
where the coefficients are given by
 \begin{eqnarray}
&C^0_{A_\chi} &= -\frac{6g^2}{p^2} \slashed{p} \\
&C^0_{A_W} &= \frac{6g^2}{p^2} \slashed{p}\\
&C^0_{B_{\chi W}} &= \frac{g^2}{\hat{m}_W^2p^2} \left( p^2+\hat{M}^2-\hat{m}_W^2   \right)\slashed{p} - 24g^2\hat{M}\\
&C^0_0 &= \left(-6g^2 + \delta_Z\right) \slashed{p}  + \left(12g^2+\delta_M\right)\hat{M}.
\end{eqnarray}

\subsubsection{Charged component}\label{app:mdm_c}

The self energy of the charged component, $\mychi^+$, is
\begin{align}
\begin{split}
\kappa^{-1}\Sigma^+(p^2) = \, & C^+_{A_\chi}\,A(\hat{M})  + C^+_{A_W}\,A(\hat{m}_W)   + C^+_{A_Z}\,A(\hat{m}_Z)      \\ &  +  C^+_{B_{\chi\gamma}}\,B(\hat{M},0)     + C^+_{B_{\chi W}}\,B(\hat{M},\hat{m}_W) + C^+_{B_{\chi Z}}\,B(\hat{M},\hat{m}_Z)    + C^+_0,
\end{split}
\end{align}
where the coefficients are given by
\begin{eqnarray}
&C^+_{A_{\chi}}  & = -\frac{6g^2}{p^2} \slashed{p} \\
&C^+_{A_W}  & =  \frac{5g^2}{p^2} \slashed{p} \\
&C^+_{A_Z}  & =  \frac{g^2\cos^2(\theta_W)}{p^2} \slashed{p}   \\
&C^+_{A_{\gamma}}  & =   \frac{g^2\sin^2(\theta_W)}{p^2} \slashed{p}  \\
&C^+_{B_{\chi W}}  & = \frac{5g^2}{p^2}\left( p^2+\hat{M}^2-\hat{m}_W^2 \right) \slashed{p} - 20g^2\hat{M}  \\
&C^+_{B_{\chi\gamma}}  & =  \frac{\sin^2(\theta_W)g^2}{p^2}\left( p^2+\hat{M}^2\right) \slashed{p} - 4g^2\hat{M} \sin^2(\theta_W) \\
&C^+_{B_{\chi Z}}  & =   \frac{\cos^2(\theta_W)g^2}{p^2}\left( p^2+\hat{M}^2-\hat{m}_Z^2 \right) \slashed{p} - 4g^2\hat{M} \cos^2(\theta_W) \\
&C^+_0   & =  \left( -6g^2  +\delta_{\chi,Z}\right)\slashed{p} + \left(12g^2 + \delta_{\chi,M}\right)\hat{M}.
\end{eqnarray}

\subsubsection{Doubly charged component}\label{app:mdm_cc}

The self energy of the doubly charged component, $\mychi^{++}$, is
\begin{align}
\begin{split}
\kappa^{-1}\Sigma^{++}(p^2) = \, & C^{++}_{A_\chi}\,A(\hat{M})  + C^{++}_{A_W}\,A(\hat{m}_W)   + C^{++}_{A_Z}\,A(\hat{m}_Z)      \\ &  +  C^{++}_{B_{\chi\gamma}}\,B(\hat{M},0)     + C^{++}_{B_{\chi W}}\,B(\hat{M},\hat{m}_W) + C^{++}_{B_{\chi Z}}\,B(\hat{M},\hat{m}_Z)    + C^{++}_0,
\end{split}
\end{align}
where the coefficients are given by
\begin{eqnarray}
&C^{++}_{A_{\chi}}  & = -\frac{6g^2}{p^2} \slashed{p} \\
&C^{++}_{A_W}  & =  \frac{2g^2}{p^2} \slashed{p} \\
&C^{++}_{A_Z}  & =  \frac{4g^2\cos^2(\theta_W)}{p^2} \slashed{p}   \\
&C^{++}_{A_{\gamma}}  & =   \frac{4g^2\sin^2(\theta_W)}{p^2} \slashed{p}  \\
&C^{++}_{B_{\chi W}}  & = \frac{2g^2}{p^2}\left( p^2+\hat{M}^2-\hat{m}_W^2 \right) \slashed{p} - 8g^2\hat{M}  \\
&C^{++}_{B_{\chi\gamma}}  & =  \frac{4\sin^2(\theta_W)g^2}{p^2}\left( p^2+\hat{M}^2\right) \slashed{p} - 16g^2\hat{M} \sin^2(\theta_W) \\
&C^{++}_{B_{\chi Z}}  & =   \frac{4\cos^2(\theta_W)g^2}{p^2}\left( p^2+\hat{M}^2-\hat{m}_Z^2 \right) \slashed{p} - 16g^2\hat{M} \cos^2(\theta_W) \\
&C^{++}_0   & =  \left( -6g^2  +\delta_{\chi,Z}\right)\slashed{p} + \left(12g^2 + \delta_{\chi,M}\right)\hat{M}.
\end{eqnarray}

\subsubsection{Gauge bosons}\label{app:mdm_vv}

The contributions from the MDM quintuplet to the gauge bosons self energies are
\begin{eqnarray}
&\Pi_{ZZ,\chi\chi} &= \frac{5e^2\cot^2(\theta_W)}{36\pi^2}\Pi(\hat{M})\\
&\Pi_{\gamma\gamma,\chi\chi} &= \frac{5e^2}{36\pi^2}\Pi(\hat{M})\\
&\Pi_{WW,\chi\chi} &= \frac{5g^2}{36\pi^2}\Pi(\hat{M})\\
&\Pi_{Z\gamma,\chi\chi} &= \frac{5e^2\cot^2(\theta_W)}{36\pi^2}\Pi(\hat{M}),
\end{eqnarray}
where $\Pi$ is as given in Eq.\ \ref{eqn:pi}.

\subsubsection{Counter-term couplings}\label{app:mdm_ct}

The counter-terms $\delta_{\chi,Z}$ and $\delta_{\chi,M}$ are given by
\begin{eqnarray}
\delta_{\chi, Z}&=&12g^2\Delta,\\
\delta_{\chi,M}&=&-48g^2\Delta.
\end{eqnarray}

Additional counter-terms for the gauge-multiplet three-point vertices are required to control divergences in the two-loop self energies.  They are
\begin{eqnarray}
\frac{\delta_{\cn\cp W^+} }{\sqrt{3}}&=& \frac{\delta_{\chipp\cp W^+} }{ \sqrt{2} } = -\frac{g^3}{\pi^2}\Delta\\
\frac{\delta_{\chipp\chipp \gamma}}{2\sin(\theta_W)} &=&\frac{\delta_{\chipp\chipp Z} }{2\cos(\theta_W)}=\frac{\delta_{\cp\cp \gamma} }{\sin(\theta_W)}= \frac{\delta_{\cp\cp Z}}{\cos(\theta_W)} = \frac{g^3}{\pi^2}\Delta.
\end{eqnarray}
We determine these terms by demanding that the two-loop self energy be free of UV divergences (i.e.\ free of any poles in $\epsilon$ or $\epsilon^2$).

The gauge boson counter-term couplings are
\begin{eqnarray}
&\delta_{Z,WW}  &=  -\frac{15g^2}{32\pi^2}\Delta\\\
&\delta_{M,WW} &= \frac{g}{32\pi^2} \left[-\sum_{i}c_im_i^2 + 15\hat{m}_W^2-2\hat{m}_Z^2\cos(2\theta_W)\right]\Delta\\\
&\delta_{Z,ZZ}  &=  -\frac{g^2}{96\pi^2}\left[43\cos(2\theta_W)+41\sec^2\theta_W-39\right]\Delta\\\
&\delta_{M,ZZ} &= -\frac{g\sec^2(\theta_W)}{96\pi^2} \left[  3\sum_i c_i m_i + \hat{m}_W^2\left(70-47\sec^2\theta_W+62\cos^2\theta_W\right)  \right]\Delta\\\
&\delta_{Z,\gamma\gamma} & =  -\frac{43g^2\sin^2\theta_W}{48\pi^2}\Delta\\\
&\delta_{M,\gamma\gamma} & = 0\\
&\delta_{Z,Z\gamma} & =  \frac{g^2}{96\pi^2}\left(41\tan\theta_W-43\sin2\theta_W\right) \Delta\\\
&\delta_{M,Z\gamma} & =   -\frac{g^2\hat{m}_Z^2}{8\pi^2}\sin\theta_W\cos\theta_W\Delta,
\end{eqnarray}
where the summation is over all SM quarks and leptons given in Eq.\ \ref{eqn:mi}.
\end{widetext}

\bibliography{../library}{}

\end{document}